\documentclass[a4paper]{article}

\usepackage[english]{babel}
\usepackage[utf8x]{inputenc}
\usepackage[T1]{fontenc}
\usepackage{parskip}
\usepackage{comment}
\usepackage{placeins}

\usepackage[a4paper,top=3cm,bottom=2cm,left=3cm,right=3cm,marginparwidth=1.75cm]{geometry}

\usepackage{amsmath}
\usepackage{graphicx}
\usepackage[colorinlistoftodos]{todonotes}
\usepackage[colorlinks=true, allcolors=blue]{hyperref}
\usepackage{authblk}
\usepackage{caption,subcaption}
\usepackage{rotating}
\usepackage{xspace}
\usepackage{mathtools}
\usepackage{booktabs}
\usepackage{textcomp}

\graphicspath{{./images/}}

\newcommand{\arXiv}{\texttt{arXiv}\xspace}
\newcommand{\MAG}{\texttt{MAG}\xspace}
\newcommand{\GRID}{\texttt{GRID}\xspace}
\newcommand{\cb}{\texttt{CrunchBase}\xspace}
\newcommand{\ce}{\texttt{CorEx}\xspace}

\title{Deep learning, deep change? Mapping the development of the Artificial Intelligence General Purpose Technology}

\author{J. Klinger}
\author{J. Mateos-Garcia}
\author{K. Stathoulopoulos}
\affil{Nesta, 58 Victoria Embankment, London, EC4Y 0DS, United Kingdom}
\date{}

\begin{document}
\maketitle

\begin{abstract}
  General Purpose Technologies (GPTs) that can be applied in many industries are an important driver of economic growth and national and regional competitiveness. In spite of this, the geography of their development and diffusion has not received significant attention in the literature. We address this with an analysis of Deep Learning (DL), a core technique in Artificial Intelligence (AI) increasingly being recognized as the latest GPT. We identify DL papers in a novel dataset from ArXiv, a popular preprints website, and use CrunchBase, a technology business directory to measure industrial capabilities related to it. After showing that DL conforms with the definition of a GPT, having experienced rapid growth and diffusion into new fields where it has generated an impact, we  describe changes in its geography. Our analysis shows China's rise in AI rankings and relative decline in several European countries. We also find that initial volatility in the geography of DL has been followed by consolidation, suggesting that the window of opportunity for new entrants might be closing down as new DL research hubs become dominant. Finally, we study the regional drivers of DL clustering. We find that competitive DL clusters tend to be based in regions combining research and industrial activities related to it. This could be because GPT developers and adopters located close to each other can collaborate and share knowledge more easily, thus overcoming coordination failures in GPT deployment. Our analysis also reveals a Chinese comparative advantage in DL after we control for other explanatory factors, perhaps underscoring the importance of access to data and supportive policies for the successful development of this complex, `omni-use'  technology.
\end{abstract}

\tableofcontents

\section{Introduction}
  What do the steam engine, the electric motor and the microprocessor have in common? They are all powerful General Purpose Technologies (GPTs) that can be applied in multiple sectors creating waves of change that ripple across the economy \cite{bresnahan1995general}.  It is not a coincidence that economic eras are often named after their `core' GPTs: the Steam Age, the Age of Electricity, the Information Revolution and today, a `Second Machine Age' driven by advances in Artificial Intelligence (AI) \cite{brynjolfsson2017artificial,mokyr2002gifts}.
    
    The emergence of a GPT can also change the economic fortunes of nations and regions: it is hard to disentangle Britain's ascendancy from the steam engine, or the USA's from electrification and the combustion engine. The arrival of microelectronics and the Internet shifted economic power from the East Coast of the US to Silicon Valley in the West. Today,  the rhetoric of an AI `global race' implies that those countries that develop strong AI industries will be able to dominate more markets and industries. Governments across the world are responding with national strategies to grow their AI sectors \cite{cave2018ai} .
    
  But where do GPTs such as AI appear and why, and how do they transform geographies of innovation and production? We still lack good answers to these questions. Although economic geographers and regional scientists have studied disruptive GPT-like innovations that create new opportunities for countries and regions, they rarely consider their links with the rest of the economy \cite{audretsch1996innovative,scott2003regions}. Yet it is precisely this connectivity that defines GPTs, and could also explain where they emerge, and their geographical impact \cite{bresnahan2010reallocating}.
   
   In this paper we seek to address this gap in the literature with an analysis of the geography of Deep Learning (DL) research, one of the technologies driving recent advances in AI systems that are increasingly being recognized  as the latest GPT \cite{agrawal_economic_2018,cockburn2018impact,furman2018ai}. We consider how the geography of DL has evolved since its emergence in the early 2010s, and study its link with local research and industrial capabilities. Our analysis draws on the literature on technological discontinuities and a recent body of research on economic complexity and related diversity that looks at how the industrial and knowledge composition of regions and countries drive their diversification into new products and technologies \cite{hidalgo2009building,frenken2007related}. In doing this, we provide new evidence about the geography of AI research, a question of great interest for policymakers.
   
  Our analysis draws on a novel combination of data sources and methods: we obtain our principal dataset from \arXiv, a preprints site widely used by scientists and engineers, and identify DL papers in its computer science section with \texttt{CorEX}, an information theory algorithm that can detect clusters of related words in corpora of text. We also use data from \cb, a technology business directory, to identify and map industries that are related to DL and might spur its development. Our data analysis pipeline illustrates the opportunities that novel data science methods create for the analysis of emerging technologies such as DL.\footnote{The code we have used in our analysis is available for review in \texttt{https://github.com/nestauk/arxiv\_ai}.}

  The rest of this section reviews relevant literatures in economics, economic geography and AI. Section \ref{sec:data} describes how we collected and enriched the data and classified papers from \arXiv computer science corpus into the DL category. Section \ref{sec:analysis} presents our findings in three steps. First, we consider whether DL displays three defining features of a GPT (\textit{rapid growth, rapid diffusion into new fields, and impact in new fields}). Second, we study the geographical aspects of its diffusion. Third, we model the link between regional specialization in DL research and activity in related knowledge and industrial bases. Section \ref{sec:conclusion} discusses the findings and its limitations, and outlines issues for further research. 

\subsection{General Purpose Technologies as engines of growth}
\label{sec:ai_gpt}
  GPTs are technologies or clusters of related technologies \textit{`characterized by the potential for pervasive use in a wide range of sectors and by their technological dynamism'} \cite{bresnahan1995general,hall2004uncovering}. They enable productivity improvements in multiple industries by automating or greatly improving the efficiency of key production tasks such as the use of energy for work, or the transfer and processing of information. The steam engine replaced human, animal and natural motor power in mining, textiles and transport \cite{mokyr2002gifts}. Electricity cheaply illuminated homes and workplaces, and the combustion engine detached energy from a fixed grid, making production and transport more flexible \cite{david1990dynamo}. Micro-electronics transformed the speed and scale of computation across the economy. 

  If we imagine the technology system as a network of ideas being constantly recombined, then we will find GPTs sitting near its center \cite{arthur2009nature}. GPTs induce cascades of complementary innovations in the sectors that deploy them, some of which may also be widely applicable. For example, Information and Communication Technologies (ICTs) based on cheap microchips gave birth to the video-games industry, which subsequently spurred the development of Graphical Processing Units (GPUs) now used for parallel processing of information in other sectors. This exploration of new GPT opportunities requires trial-and-error and can take time. For example, US factories did not start to realize the benefits of electric power until they reorganized their layout to harness the flexibility of small electric motors, decades after the introduction of electricity \cite{david1990dynamo}.
  
  The networked nature of GPTs creates the risk of \textit{coordination failures}  in its deployment: rapid change makes their evolution hard to predict, and might encourage a `wait-and-see' strategy among potential adopters and providers of complementary skills, infrastructures and standards. This can hinder the exploration of the new opportunities the GPT offers, and delay or halt follow-on innovations \cite{NBERw4854}.

\subsection{Towards a geography of GPTs}
\label{sec:gpt_geo}
  When they arrive, GPTs transform the economic conditions and production processes of many industries. Consider for example the changes brought about by the advent of steam to textiles and transport in Britain, or more recently, the impact of the Internet in media or retail. Since industries tend to cluster in specific locations to access dense talent pools, reduce transaction costs and learn from each other, the impact of GPTs will also be unequally distributed in space \cite{porter1998clusters}. If a GPT is `competence-destroying' for an industry (that is, if it eliminates previous sources of comparative advantage like the Internet did with control over physical distribution channels in the music industry), then those locations where the industry concentrates will experience a negative shock. At the same time, a GPT can create windows of opportunity to enter a sector, like the Internet did with new media clusters. 
 
  Economic geographers have studied similar discontinuities through the lens of the product life-cycle. The idea is that the technologies used by an industry follow a trajectory with distinct phases, and that each of these phases has a different geography. Early in the life-cycle, when a new market or technological opportunity is revealed, there is a phase of experimentation when entrepreneurs explore different designs to harness this opportunity \cite{anderson1990,abernathy1978}\footnote{(the beginning of the automobile industry is a paradigmatic example of this phase, with inventors and entrepreneurs exploring in parallel various energy sources for the automobile, from the combustion engine to electrical and steam-powered motors, \cite{klepper1996entry})}. In this phase of the product life-cycle, there is uncertainty about the technologies and capabilities required to succeed in the market, lowering barriers to entry for new entrepreneurs and regions \cite{scott2003}. Eventually, this experimentation yields a standard or dominant design and the industry moves from product to process innovation. Economies of scale become more important, leading to industrial and geographical consolidation.\footnote{At the same time, there might be some dislocation of activity as standardized parts of the production process are outsourced or off-shored to other locations with cheaper costs.} We would expect something similar to happen when a GPT arrives, with an initial phase of geographical volatility when new entrants come into the market, followed by a shake-out and increasing concentration once a dominant design is established.
  
  What factors determine whether a region is able to enter and successfully compete in the development of the GPT in the first place? A growing body of literature on \textit{Economic Complexity} and \textit{Economic Relatedness} suggests that a region's ability to enter a new market or technology depends on the presence of related capabilities  that can be re-purposed or recombined to explore new opportunities. This is referred to as the \textit{Principle of Relatedness} \cite{hidalgo2009building,frenken2007related,klepper1996entry,delgado2018principle}. Building on this idea, GPTs that can be applied in multiple industries could benefit from the co-location of R\&D sectors that develop the technology and industrial sectors where it can be applied. Proximity between developers, adopters and suppliers of skills and infrastructure facilitates communication and reduces the risk of coordination failures, improving the prospects for GPT deployment and helping the location gain a comparative advantage in the technology \cite{boschma2005proximity}. 
  
\subsection{Empirical setting: Artificial Intelligence and Deep Learning}
\label{sec:ml_ai}
  Having discussed the concept of GPTs, we now turn our attention to the empirical setting for our analysis: Artificial Intelligence, and more specifically the Deep Learning techniques underpinning it.

  Artificial Intelligence (AI) systems have been defined as \textit{`self-training structures of Machine Learning predictors that automate and accelerate human tasks'} \cite{taddy2018technological}. In turn Machine Learning (ML) is \textit{`the field that thinks about how to automatically build robust predictions from complex data'} \cite{taddy2018technological}. ML emerged in the 1970s in response to the failure of rule-based approaches where human experts hard-coded knowledge in Artificial Intelligence systems \cite{markoff2016}. ML's approach is to instead develop algorithms that can recognize patterns in labeled data with less need for human intervention, and use the resulting models to make predictions about new observations. Economic analyses of AI focus on its ability to reduce the costs of prediction, an important task in many industries \cite{agrawal2018prediction}. 

  Deep Learning (DL) is a new ML technique that processes large and complex datasets it through networks of synthetic neurons where subsequent layers learn increasingly abstract representations of the data that eventually become an input into prediction \cite{goodfellow2016}. Although the neural network literature goes back to the 1950s, this approach only became feasible in recent years thanks to the availability of large, labeled datasets from the web, and powerful GPUs. Since the early 2010s, Deep Learning has been proven to be \textit{`unreasonably effective'} in many applications, from image and video recognition to translation and gaming, fueling a surge of interest and investment in AI \cite{karpathy2015}. 

  Ultimately, AI researchers strive for generality: developing algorithms that can transfer their predictive prowess across domains, and respond effectively to new situations. Sustained progress towards that goal has led a growing number of economists to declare DL-driven AI a new GPT that will revolutionize the economy \cite{brynjolfsson2017artificial}. DL also represents an \textit{`invention in the methods of invention'} that could transform how new ideas are discovered, improving productivity of R\&D in fields such as drug discovery, genomics or material sciences \cite{cockburn2018impact,agrawal2018finding}. Publication, patenting and venture capital trends support this view, with rapid growth in DL activity and diffusion into other disciplines and industries \cite{cockburn2018impact}. 
  
  The GPT nature of AI would also explain stagnant productivity growth despite rapid technological progress: businesses still need to reorganize their operations \cite{brynjolfsson2017artificial}, the education system needs to address skills shortages, and suitable digital and regulatory infrastructures have to be developed to create value from AI-driven growth.
  
  What about the geography if AI? A recent review of its international trade aspects argues that the localized nature of AI knowledge spillovers (the fact that organizations need to be based in the locations where investments on R\&D take place to benefit from them) could justify national policies to support its development \cite{goldfarb2018ai}. Governments across the world appear to share this view, and many have announced national strategies to compete in the `AI global race'. There is a growing belief that China, with its large STEM workforce, powerful Internet platforms and vast amounts of data is `winning' this race \cite{williams_why_2018}. Meanwhile, European researchers and policymakers fear that the EU falling behind for lack of talent and leading AI-driven businesses  \cite{editor_scientists_2018}. These perceptions imply that AI GPT is disrupting the geography of digital production and innovation. As AI researcher Andrew Ng points out  \textit{`Since AI changes the foundation of many technology systems - everything ranging from web search to autonomous driving to customer service chatbots - it also gives many countries the opportunity to `leapfrog' the incumbents in some application areas'} \cite{ai_index_artificial_2017}. In the rest of this paper, we monitor these geographical changes and study their drivers using a novel preprints dataset and state-of-the-art Natural Language Processing (NLP) methods.
  
\section{Data collection and classification}
\label{sec:data}

Our analysis relies on several data sources and preprocessing activities:

\begin{enumerate}
\item We combine data from \arXiv, \GRID (Global Research Identifier) and \MAG (Microsoft Academic Graph) to create a geocoded dataset of research activity in computer science disciplines where we identify DL papers with \ce, a topic modeling algorithm. We also measure the relatedness between computer science subjects based on their co-occurrence in \arXiv papers.
\item We use \cb, a business directory, to map industrial activities that might be relevant for the development of DL clusters. We measure relatedness between those industries and DL using  a machine learning model that predicts industrial sectors with company descriptions.
\end{enumerate}

We go through these two streams of data collection and classification in turn.

\subsection{Identifying and mapping DL papers in \arXiv data}
\label{subsec:DLdata}

 We generate the DL dataset for our analysis by matching three non-proprietary open data sources;
\arXiv, Microsoft Academic Graph (\MAG), and the Global Research Identifier Database (\GRID). The data 
sources are matched in the following order, according to the procedure described in Sections~\ref{subsubsec:arxiv}-~\ref{subsubsec:grid}:

\begin{center}
  \begin{math}
  	\{ \arXiv \xrightarrow{\text{matched to}} \MAG \} \xrightarrow{\text{matched to}} \GRID
  \end{math}
\end{center}

 By following this pipeline of data collection, we create a dataset with the features described in Table~\ref{tab:datasources} for further processing as described in Section~\ref{subsubsec:topic}.

\begin{table}[ht]
  \centering
  \begin{tabular}{ l  l  p{60mm} }
    \bf{Feature} & \bf{Data source} & \bf{Comments} \\
    \hline
    Article title & \arXiv & Assured to be consistent with \MAG title after matching procedure.\\
    Article abstract text & \arXiv & To be used for topic modeling (Section~\ref{subsubsec:topic}).\\
    Subject classification & \arXiv & Assigned by the author. \\
    Is article published? & \MAG & Always true, as implicitly assured by match to \MAG. \\
    Publication date & \MAG & Publication date in MAG, rather than \arXiv submission date. \\
    Citation count & \MAG & Used for cross-check by selecting 'high quality' publications (Section~\ref{sec:analysis}).\\

    Institute affiliation (all authors) & \MAG & This replaces the potentially incomplete set of authors from \arXiv. \\
    Institute location & \GRID & 
  \end{tabular}
  \caption{Features extracted in the data collection procedure.}  
  \label{tab:datasources}
\end{table}

  \subsubsection{\arXiv}
  \label{subsubsec:arxiv}
  \arXiv is a `real-time' open archive of academic preprints widely used by researchers in quantitative, physical and computational science fields. Data from each of over 1.3~million papers can be accessed programmatically via the \arXiv API. As \arXiv papers are self-registered, we ensure that papers are not simply `junk' articles by requiring that all papers are matched to a journal publication or conference proceeding, as presented in Section~\ref{subsubsec:mak}. We also have anecdotal evidence that the archive contains many high quality papers, since a short study of conference proceeding from the prestigious AI Conference on Neural Information Processing Systems in 2017 reveals that over 55\% of these were published on \arXiv. 
  
  Is \arXiv a suitable data source for the analysis of industrial R\&D? We believe that this is the case. The AI research community has a strong culture of openness in its publication of research findings, software and benchmark datasets, which are perceived as a way to attract scientific talent \cite{bostrom2017strategic}. Some of the most active DL institutions in our corpus include corporations such as Google, Microsoft, IBM, Baidu or Huawei. 
 
  From the initial set of over 1.3~million papers, approximately 134,000 have been selected for analysis as they fall under the broad category of `Computer Science' (\texttt{cs}) or the specific category of `Statistics - Machine Learning' (\texttt{stat.ML}).
  
  \subsubsection{Microsoft Academic Graph (\MAG)}
  \label{subsubsec:mak}
  
  Microsoft Academic Graph (\MAG) is an open API offering access to 140~million academic papers and documents compiled by Microsoft and available as part of its `Cognitive Services'. For the purpose of this paper, \MAG helps to ensure that article retrieved from \arXiv have been published in a journal or conference proceeding, as well as providing citation counts, publication date and author affiliations. The matching of the \arXiv dataset described in Section~\ref{subsubsec:arxiv} is performed in two steps.

  We begin by matching the publication title from \arXiv to the \MAG database. The database can be queried by paper title, although fuzzy-matching\footnote{`Fuzzy-matching' refers to the process of finding a likely match for a set of text (such as a word or sentence) amongst a choice of texts. A naive example would be comparing the ratio of the number of characters between texts, and identifying the texts with the highest ratio as a match.} or near-matches are not possible with this service. Furthermore, since paper titles in \MAG have been preprocessed, one is required to apply a similar preprocessing prior to querying the \MAG database. There is no public formula for achieving this, so we explicitly describe the following steps to emulate the \MAG preprocessing:
  
  \begin{enumerate}
    \item Identify any `foreign' characters (for example, Greek or accented letters) as non-symbolic;
    \item Replace all symbolic characters with spaces; and
    \item Ensure no more than one space separates characters.
  \end{enumerate}

  This procedure leads to a match rate of 90\%, for the set of \arXiv articles used in this paper. We speculate that papers could be missing for several reasons: the titles on \arXiv could significantly different from those on \MAG; the latter procedure may be insufficient for some titles; the \arXiv paper may not be published in a journal; and \MAG may not otherwise contain the publication. It may be possible to recuperate some of these papers, however this is currently not a limiting factor in our analysis.

  \subsubsection{Global Research Identifier Database (\GRID)}
  \label{subsubsec:grid}
  
  We use the Global Research Identifier Database (\GRID) to enrich the dataset with geographical information, specifically a latitude and longitude coordinate for each affiliation that we can then geocode into countries and regions.\footnote{We do this with a point-in-polygon approach using boundary (shapefile) data from the \href{https://www.naturalearthdata.com/}{Natural Earth} public map dataset.}. The \GRID data is particularly useful since it provides institute names and aliases (for example, the institute name in foreign languages). Each institute name from \MAG is matched to the comprehensive list from \GRID as follows:

  \begin{center}
  \begin{enumerate}
      \item If there is an exact match amongst the institute names or aliases, then extract the coordinates of this match. Assign a `score' of 1 to this match (see step 3. for the definition of `score').
      \item Otherwise, check whether a match has previously been found. If so, extract the coordinates and score of this previous match.
      \item {Otherwise, find the \GRID institute name with the highest matching score, by convoluting the scores from various fuzzy-matching algorithms in the following manner:

      \begin{equation} 
          \frac{1}{\sqrt{N}} \sqrt{ \sum_{n=0}^{N} F_{n}(m_{\text{\MAG}},M_{\text{\GRID}})^{2} }
          \label{eq:makscore}
      \end{equation}

    where $N$ is the number of fuzzy-matching algorithms to use, $F_{n}$ returns a fuzzy-matching score (in the range $0 \rightarrow 1$) from the $n^{\text{th}}$ algorithm, $m_{\text{\MAG}}$ is the name from \MAG to be matched and $M_{\text{\GRID}}$ is the comprehensive list of institutes in the \GRID data.}
    \end{enumerate}

  \end{center}

  The form of Equation~\ref{eq:makscore} ensures that effect of a single poor fuzzy-matching score is to vastly reduce the preference for a given match. Therefore, good matches are defined according to Equation~\ref{eq:makscore} as having multiple good fuzzy-matching scores, as measured according to different algorithms. We use a prepackaged set of fuzzy-matching algorithms implementing the Levenshtein Distance metric~\cite{levenshtein1966}, and specifically, two algorithms applying a token-sort-ratio and a partial-ratio respectively. 

  After this stage of data matching, we are left with approximately 240,000 unique institute-publication matches with at least one computer science subject in their \arXiv categories.

\subsubsection{Topic modeling}
\label{subsubsec:topic}
  
  We analyze the abstracts in our corpus using Natural Language Processing to identify papers related to DL. This involves tokenizing the text of the abstracts and removing common stop-words, very rare words and punctuation. We lemmatize the tokens based on their part-of-speech tag, and create bi-grams and tri-grams. Documents with less than twenty tokens are removed from the sample. After these steps, there are over 168,000 features (unique `words') in the dataset. 
  
  There are different approaches to identify DL papers in this preprocessed corpus. Previous work has used a keyword-search approach based on a predefined vocabulary of terms \cite{cockburn2018impact}. Here, we follow an alternative topic modeling strategy which identifies clusters of words in the data without an initial vocabulary, and provides a score for each topic in a document, simplifying the labeling process. 
  
  More specifically, we use the Correlation Explanation (\ce~\cite{versteeg2014}) algorithm, which takes an information-theoretic approach to generate $n$ combinations of features in the data which maximally describe correlations in the dataset. Using a one-hot bag-of-words representation, we optimally find $n = 28$ topics by tuning $n$ with respect to the `total correlation' variable, as advised by the \texttt{CorEx} authors. The generated topics contain words which are sorted in terms of their contribution of each feature to total correlation. We assign a score $S^j$ for each topic $j$ (containing $N^j$ words $w_i$ with topic weights $T^{j}_{i}$) to each document $W$ such that:
  \begin{gather}
      S^j = \sum_{i=0}^{N^{j}} T^{j}_{i}\delta(w_i,W) 
  \end{gather} 
  where:
  \begin{gather}  
      \delta(w_i,W) = 
      	\begin{cases}
          1 & \text{if }w_i\in W\\
          0 & \text{otherwise}
        \end{cases}
  \end{gather}
  
Topics are then assigned to each document only if the following condition is satisfied:
  \begin{gather}
  S^j \geq \gamma \ T^{j}_{\text{max}}
  \end{gather}

where $\gamma$ is a threshold parameter that we assign below, and $T^{j}_{\text{max}}$ is the maximum topic weight. The form of the above asserts that documents must contain a sufficient number of components of topics to be assigned to the topic. Clearly, a larger choice of $\gamma$ leads to a lower frequency of documents assigned to the topic whilst improving the overall recall.

  After inspecting the model outputs, we identify two topics related to DL, containing keywords such as \texttt{neural\_network}, \texttt{deep\_learning}, or \texttt{convolutional\_neural\_networks}. We label as `Deep Learning' those papers where either of these topics is present with a $\gamma$ above 0.5, giving us a set of 15,062 DL papers (11\% of the total unique papers).\footnote{We also create a more restrictive DL category containing only those papers either topic is present with a $\gamma$ above 0.5, resulting in a total of 1,604 papers. A visual inspection of a random sample of papers in both groups suggests that their outputs are similarly relevant so we opt to focus on the larger set. This is further motivated by our interest in understanding the diffusion of DL methods in various computer subjects.}. 
  
  \subsubsection{Research relatedness}
  \label{subsubsec:compprox}

  In Section \ref{subsec:drivers}, we study the link between DL specialization in a region and the presence of related research and industrial activities. We proxy research relevance using the relatedness between research subjects based on their co-occurrence in \arXiv papers.\footnote{Researchers who submit their papers to \arXiv label them with a set of relevant research categories. We focus our analysis in Computer Science (\texttt{cs}) subjects as well as those in the \texttt{stat.ML} subject.} To measure this relatedness, we calculate the cosine similarity between vectors representing the subjects that appear in different papers in the corpus. Sub-section \ref{subsec:descr} presents the results.

\subsection{Building the industrial dataset}

\subsubsection{\cb}
\label{subsubsec:cb}

  We use \cb, a commercial directory of technology companies, to measure industrial activity in a region. The version of \cb we use contains information about 257,000 organizations, including a short description of their activities, the sectors where they operate, the year when they were founded and their geographical coordinates.\footnote{As before, we geocode \cb companies using a point-in-polygon approach with boundaries from Natural Earth.} 

  Recent analyses of technology clusters in \cb suggest that it correlates well with other measures of regional technological activity, and it is increasingly being used in economics and management research \cite{dalle2017using,breschi2018portrait}. \cb presents two important advantages for our analysis: first, it has global coverage (like our \arXiv corpus) and individual organization locations, so it is easy to merge with our \arXiv data at a suitable geographical level. Second, it contains text descriptions of company activities and labels for the sectors where they operate, which we can use to generate measures of similarity between these sectors and DL papers in the \arXiv data using the strategy we describe below.

\subsubsection{Research-industry relatedness}
\label{subsubsec:indprox}
  We estimate the relatedness between industrial activities in \cb and research in \arXiv by training a supervised machine learning model that predicts the sector where a company in \cb operates based on its description\footnote{We focus on those observations with the longest and more informative descriptions, comprising around 115,000 companies. We perform a grid-search to select the best performing model, a logistic regression classifier with L1 regularization.}. 

  This model is then used for out-of-sample prediction of the \cb categories of \arXiv papers, based on the text in their abstract. Specifically, we assign categories where the prediction probability is at least 0.99.\footnote{By setting a high threshold for classification of \arXiv papers into \cb categories we seek to remove noise in the transference of the model across corpora with potential differences in their languages.} We then calculate the share of papers by \arXiv subject predicted to be in a \cb category to measure their relatedness. Subsection \ref{subsec:descr} presents the results.

\section{Analysis}
\label{sec:analysis}

\subsection{Descriptives}
\label{subsec:descr}

\subsubsection{\arXiv}
\label{subsubsec:arxiv}

  Table \ref{tab:descr} presents some descriptive statistics for papers classified as DL and the rest of the corpus. DL papers have, on average, been published more recently, they tend to contain fewer \arXiv subjects, and involve collaborations with a somewhat higher number of institutions. They also tend to receive more citations , specially  after we control for the number of years since publication. This suggests that DL is a relatively recent topic, and that DL papers are, on average, more influential than the rest.

\begin{table}[!h]
\centering
\begin{tabular}{lrr}
\toprule
dl\_cat &         dl &      non\_dl \\
\midrule
total                   &  15602 &  115587 \\
year\_average            &   2015.937 &    2013.572 \\
field\_average           &      2.798 &       2.930 \\
institute\_average       &      1.754 &       1.705 \\
citation\_average        &     25.627 &      18.003 \\
citation\_p\_year\_average &      6.182 &       2.505 \\
\bottomrule
\end{tabular}
\caption{Descriptive statistics for DL / non-DL papers in \arXiv dataset}
\label{tab:descr}
\end{table}

\begin{figure}[!h]
\includegraphics[width=\textwidth]{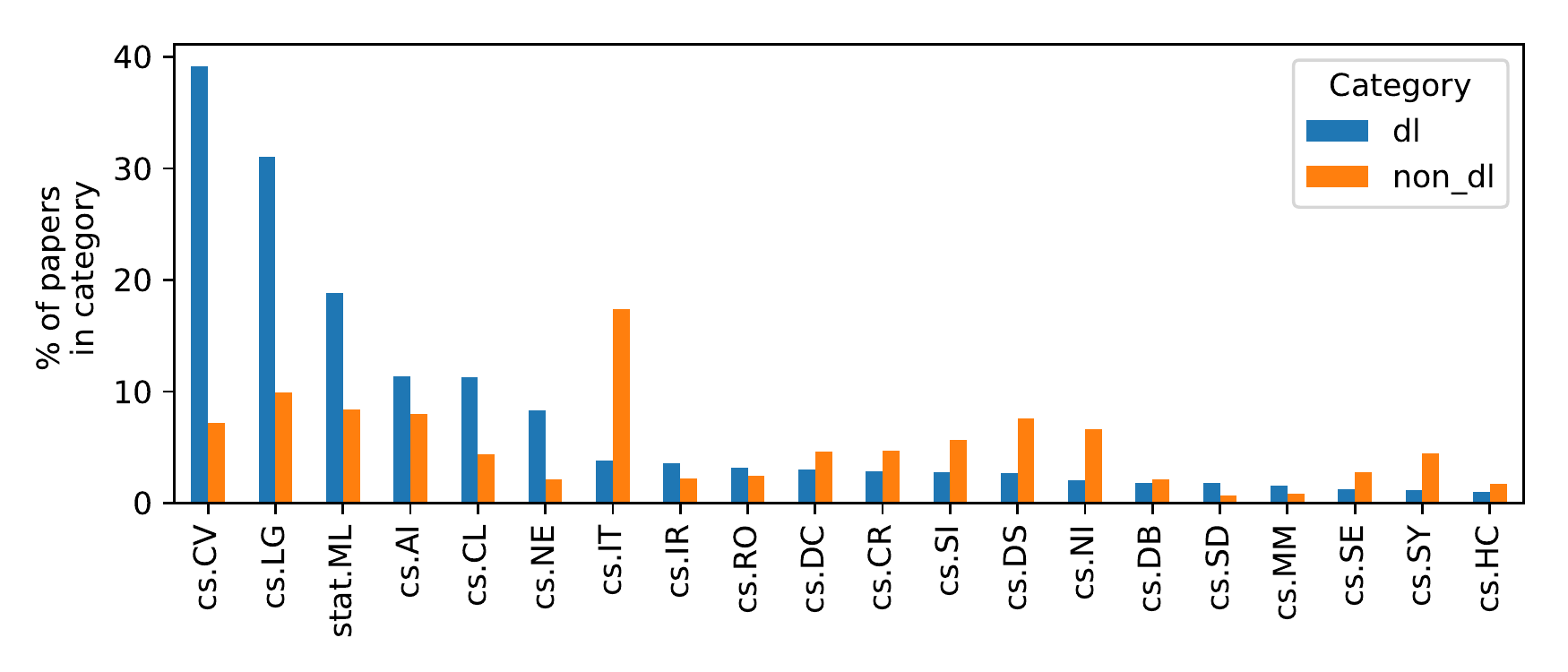}
\caption{Distribution of DL/non DL papers by \arXiv category}
\label{fig_1:cats}
\end{figure}

\begin{figure}[!h]
\includegraphics[width=\textwidth]{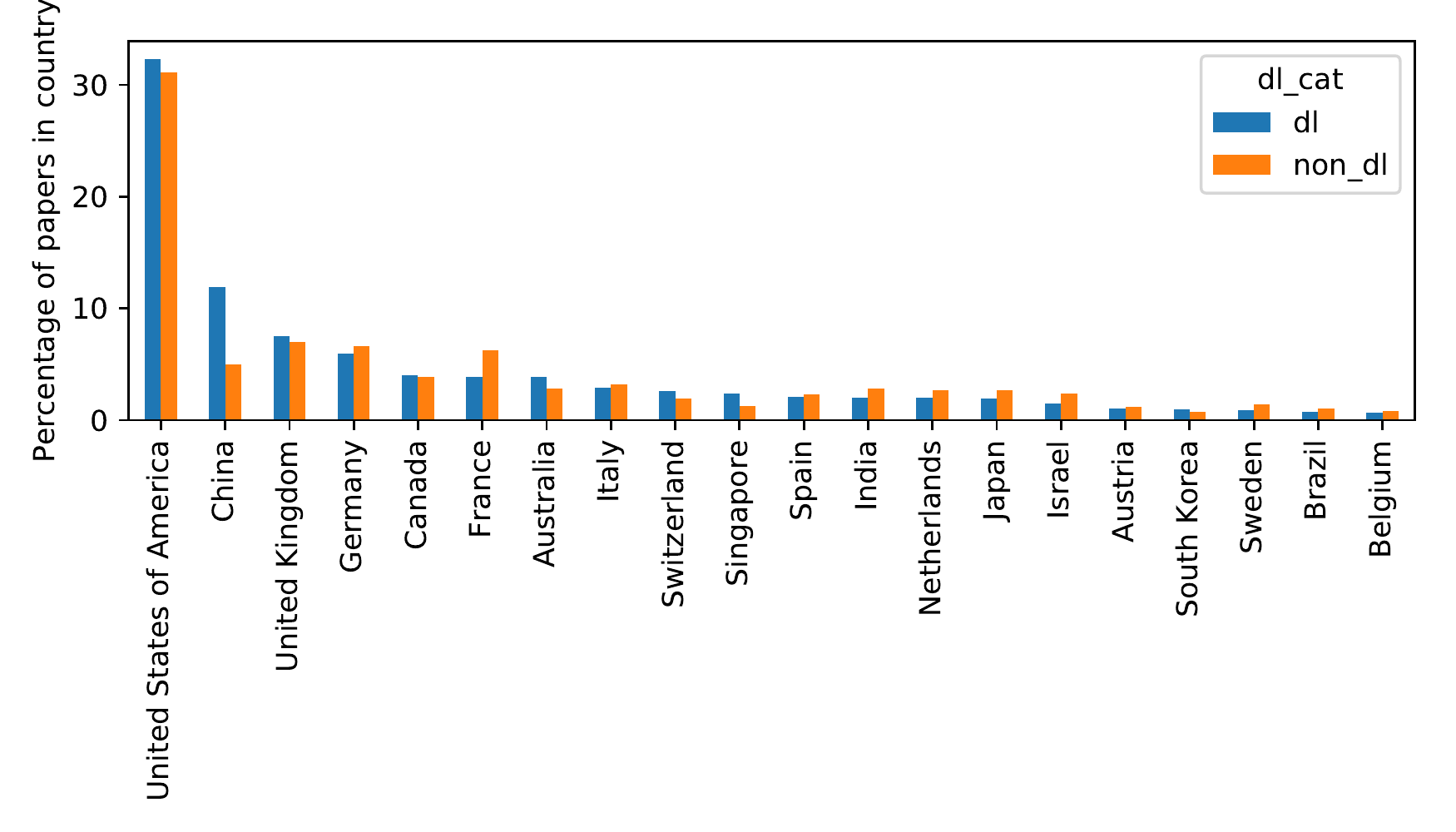}
\caption{Distribution of DL/non DL papers by country (top 20 countries)}
\label{fig_2:countr}
\end{figure}

\begin{figure}[!h]
\includegraphics[width=\textwidth]{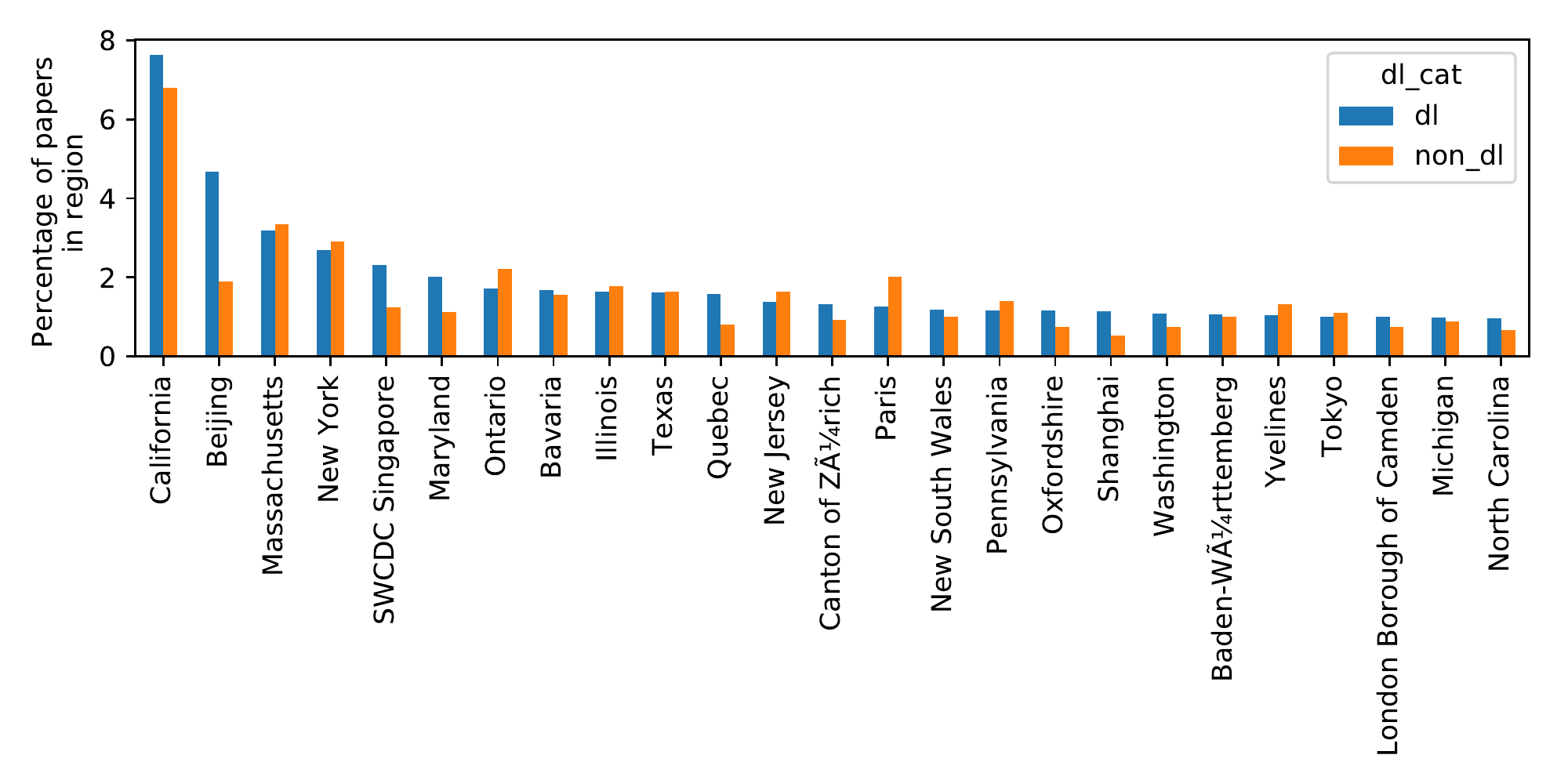}
\caption{Distribution of DL/non DL papers by region (top 35 regions)}
\label{fig_3:reg}
\end{figure}

  Figures \ref{fig_1:cats}, \ref{fig_2:countr} and \ref{fig_3:reg} present the distribution of DL and non-DL activity over \arXiv computer science subjects, countries and regions for the top categories in each variable. 

Some observations:

\begin{enumerate}
\item DL papers are highly concentrated in a small number of \arXiv subjects: Computer Vision (\texttt{cs.CV}), Computer Learning (\texttt{cs.LG}), Machine Learning (\texttt{stat.ML}), Artificial Intelligence ( \texttt{cs.AI})  and Neural Networks (\texttt{cs.NE}). The set of DL-intensive subjects includes some that rely on unstructured datasets where DL has achieved important breakthroughs, and in fields that specialize in the development of ML and AI methods.
\item The US has the biggest share of DL and non-DL papers, with around a third of all publications in both categories. China is overrepresented in DL: its share of DL papers is more than double its share of non-DL papers. By contrast, France is underrepresented in DL.
\item North American regions dominate the global rankings of DL activity. California, Massachusetts, New York, Maryland, Illinois and Texas rank highly by volume of DL activity. Ontario and Quebec in Canada also have high levels of activity, consistent with Canada's strong research base on AI. Beijing, the South West Development Corporation in Singapore, Maryland and Quebec are over-represented in DL, with substantially higher shares of activity in DL than in the rest of the corpus). Notably, only one EU region (Bavaria) appears in the top ten of global DL research in \arXiv. 
\end{enumerate}

\begin{figure}[!h]
\centering
\includegraphics[width=\textwidth]{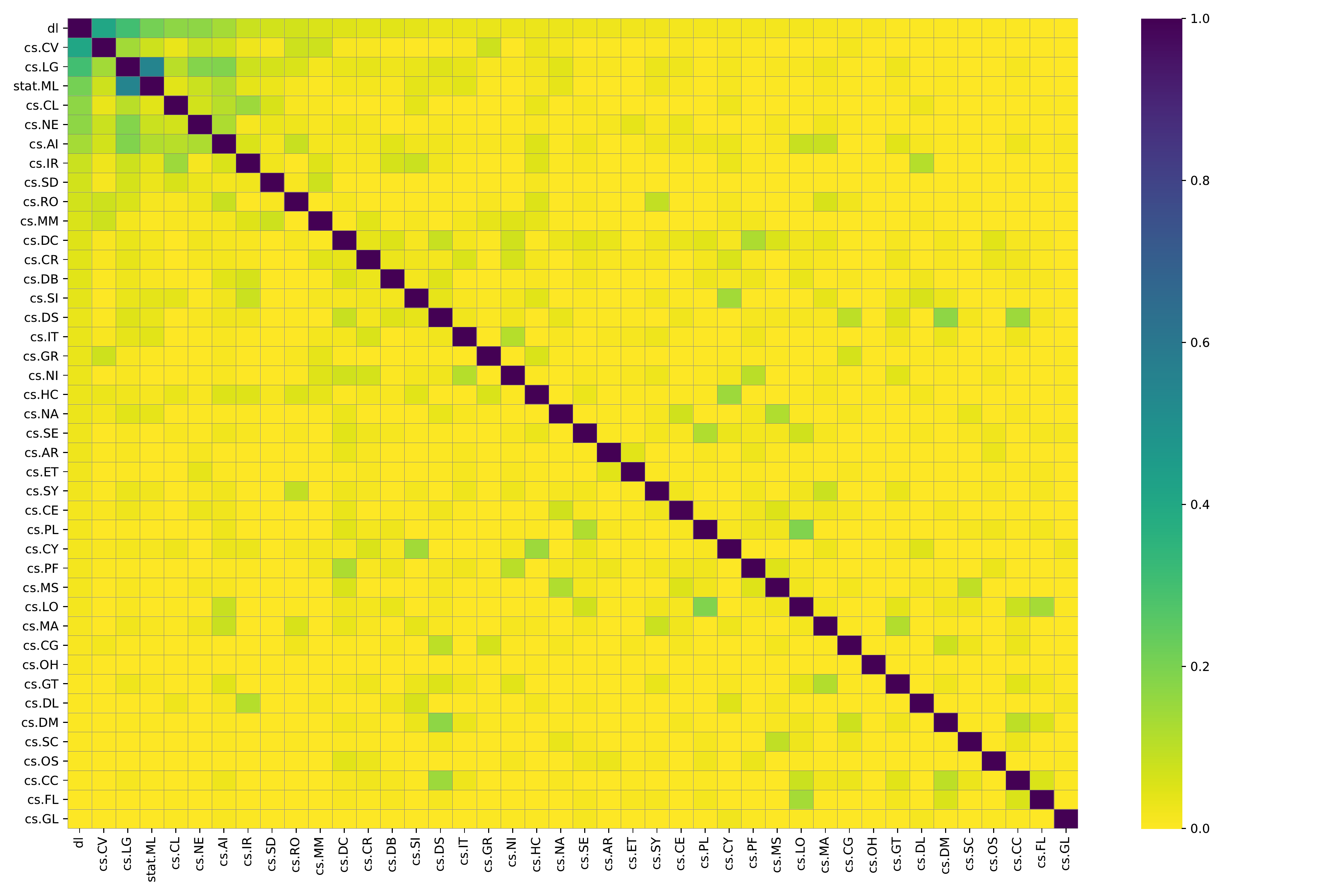}
\caption{Proximities between \arXiv subjects based on co-occurrence in papers}
\label{fig_4:arx_heat}
\end{figure}

Figure \ref{fig_4:arx_heat} displays a heatmap of the proximities between different \arXiv subjects (as well as the DL category) based on their co-occurrence on papers, sorted by their proximity to the DL category. 

Consistent with Figure \ref{fig_1:cats}, DL papers are closer to computer science subjects involving unstructured data and subjects that research ML, AI and neural networks. These subjects also tend to to co-occur with each other, forming a `cluster' of data analytics research in \arXiv. Our analysis also reveals intuitive connections between other \arXiv subjects such as Computers and Society (\texttt{cs.CY}) and Human Computer Interaction (\texttt{cs.HC}) or between Logic (\texttt{cs.LO}) and  Programming Languages (\texttt{cs.PL}), supporting the idea that these proximities are a meaningful measure of relatedness between computer science subjects in \arXiv. 

\subsubsection{\cb data}
\begin{figure}[p]
\centering
\includegraphics[width=\textwidth]{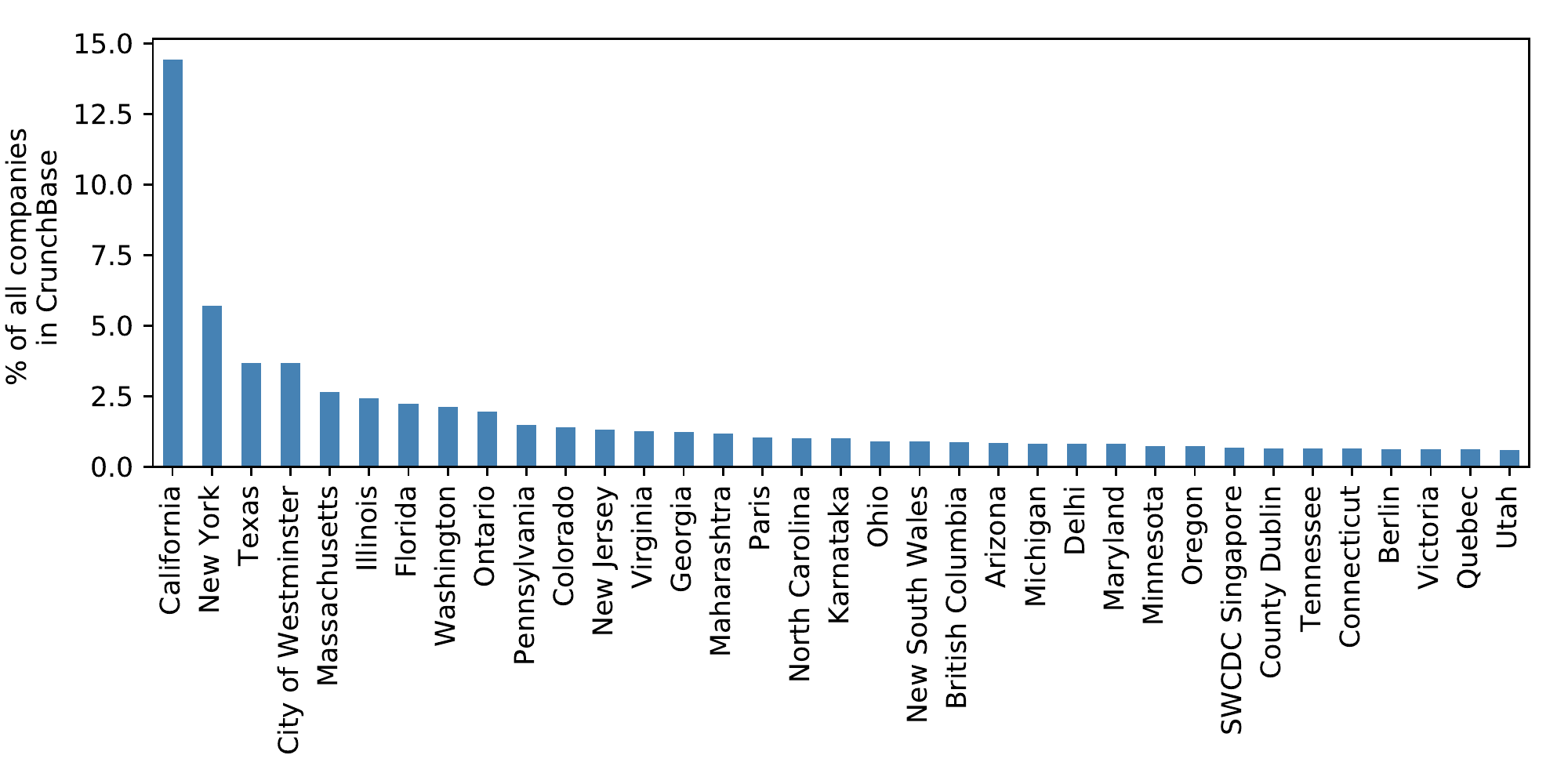}
\caption{Share of \cb activity by region}
\label{fig_5:cb_loc}
\end{figure}

Figure \ref{fig_5:cb_loc} presents the regional distribution of activity in \cb. California is again the top region by number of organizations.  Technology company activity in \cb is more concentrated than research in \arXiv (California accounted for 15\% of all activity in \cb, while it only captured 7\% of the activity in \arXiv). US States and Indian regions have a stronger presence here than they did in \arXiv. Chinese provinces are, by contrast, less visible. 

\begin{figure}[p]
\centering
\includegraphics[width=\textwidth]{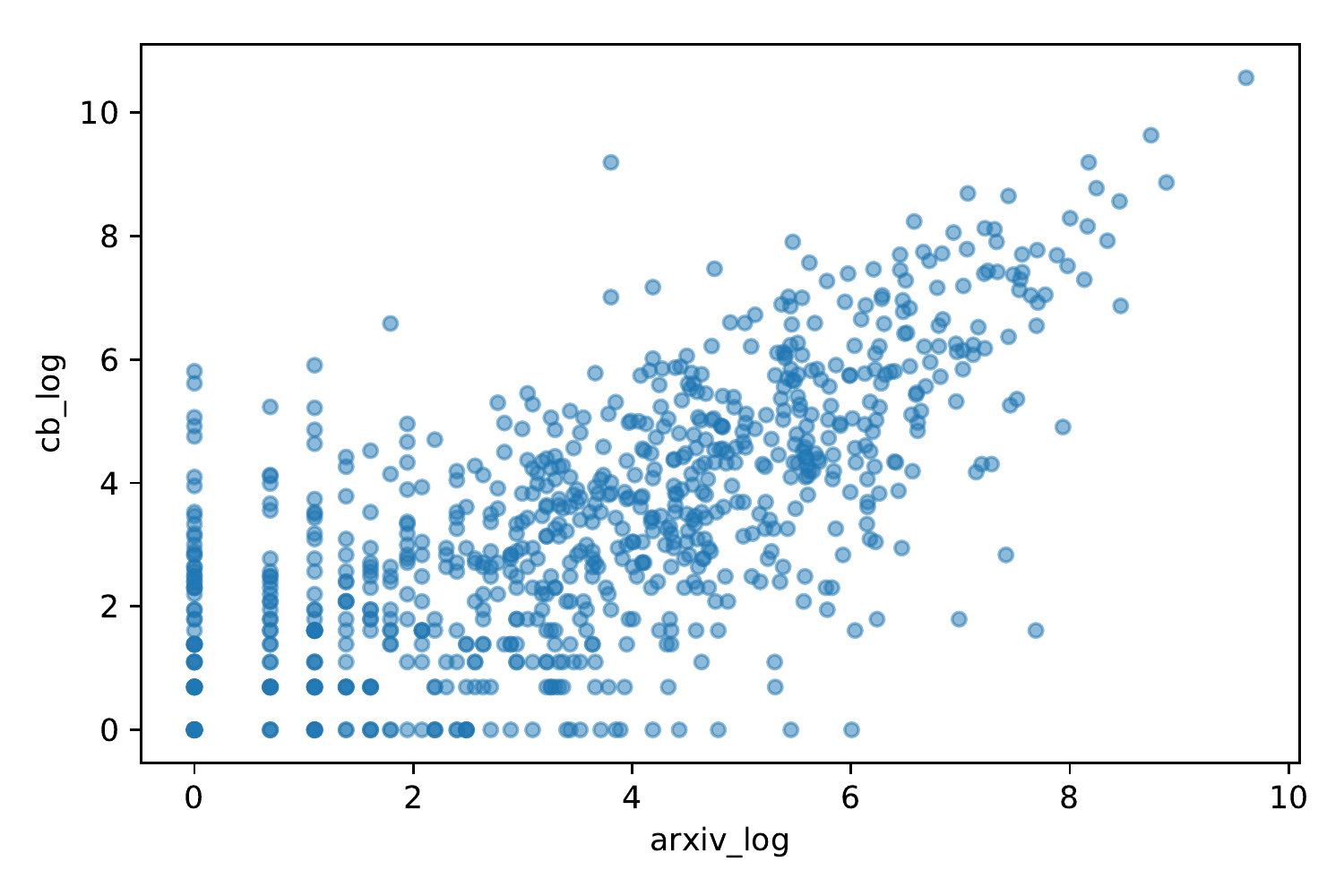}
\caption{Relationship between \arXiv and \cb activity (logged)}
\label{fig_6:arx_cb_corr}
\end{figure}

Figure \ref{fig_6:arx_cb_corr} compares levels of activity in \arXiv and \cb. Although there is a strong correlation between both datasets ($\rho$=0.67), we note some divergences. For example, there are several UK counties around London with a strong presence in \cb but low activity in \arXiv. Conversely,  some Japanese prefectures display high levels of \arXiv activity but few organizations in \cb\footnote{These results underscore the importance of triangulating our results against other data sources in future research.}.

\begin{figure}[p]
\centering
\includegraphics[width=\textwidth]{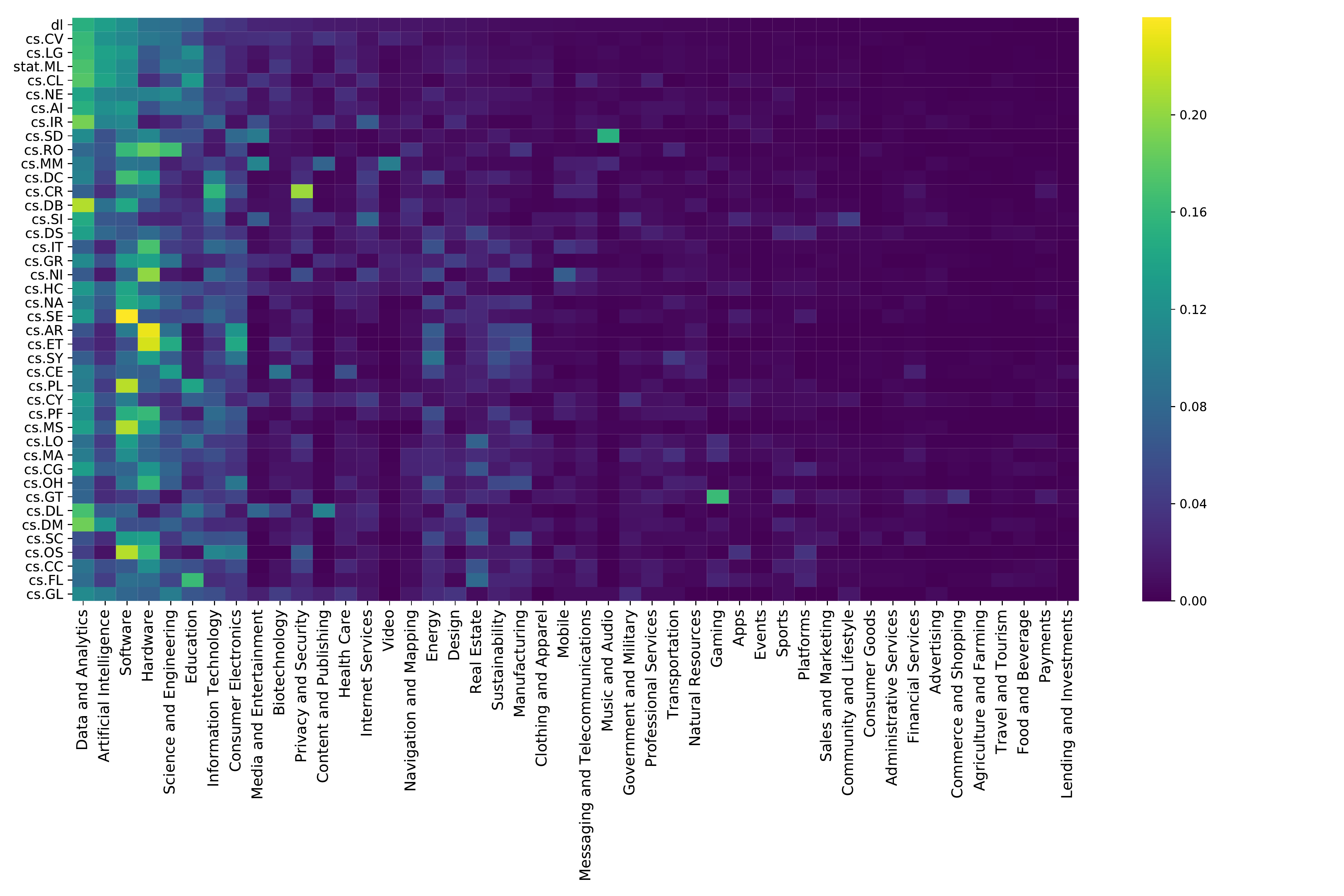}
\caption{Proximity between \arXiv disciplines and \cb sectors}
\label{fig_7:cb_sim}
\end{figure}

We end our descriptive analysis by considering the proximity between \arXiv categories (including DL) and \cb sectors based on the machine learning analysis outlined in \ref{subsubsec:cb}. The heatmap in \ref{fig_7:cb_sim} presents the share of all papers in an \arXiv subject (and DL) that were labeled in a \cb category. It shows that DL papers were classified more often in Data Analytics, Artificial Intelligences and Software \cb sectors. We also detect intuitive relations between other \arXiv categories and \cb sectors: for example, Robotics (\texttt{cs.RO}) is related to Science and Engineering, Sound (\texttt{cs.SO}) is related to Music and Audio, and Cryptography (\texttt{cs.CR}) is related to Privacy and Security. It is however worth noting that some of the similarities we identify could be linguistic rather than semantic (for example, our model detects a strong similarity between Game Theory - \texttt{cs.GT} and Gaming, which could be partly explained by their use of similar language rather than a shared knowledge base).

\subsection{GPT aspects of DL research in \arXiv}
\label{subsec:gpt_dl_analysis}

  We now move into our first question: Is DL a GPT? In answering this, we seek to ensure that our interpretation of further results is valid, and contribute to the literature on the GPT nature of AI using a new dataset and classification method \cite{cockburn2018impact}.  Previous analyses of patent data in \cite{hall2004uncovering} have looked for GPTs using patent class growth and citations, while more recently, \cite{cockburn2018impact} measure growth in DL publishing and patenting with a keyword-based approach. They also consider levels of publishing in application fields outside of Computer Science to measure the generality of DL\footnote{It is interesting to note that they classify computer vision papers and patents outside of DL. This contrasts with our finding that Computer Vision is one of the main application areas for DL, underscoring the value of unsupervised approaches for the analysis of fast moving technology fields.}. Our analysis builds on all this work.
  
  Inspired by the original definition of a GPT, we have devised the following three GPT tests for DL:
  
  \subsubsection{Rapid growth}\label{subsubsec:growth}
  The first component of the definition of a GPT is `technological dynamism', which we measure, like \cite{cockburn2018impact}, by looking at growth in activity. If DL is a GPT with broad applicability, we should see an increase in the number of DL papers in \arXiv as more researchers explore its potential.
  
\begin{figure}[p]
\includegraphics[width=\textwidth]{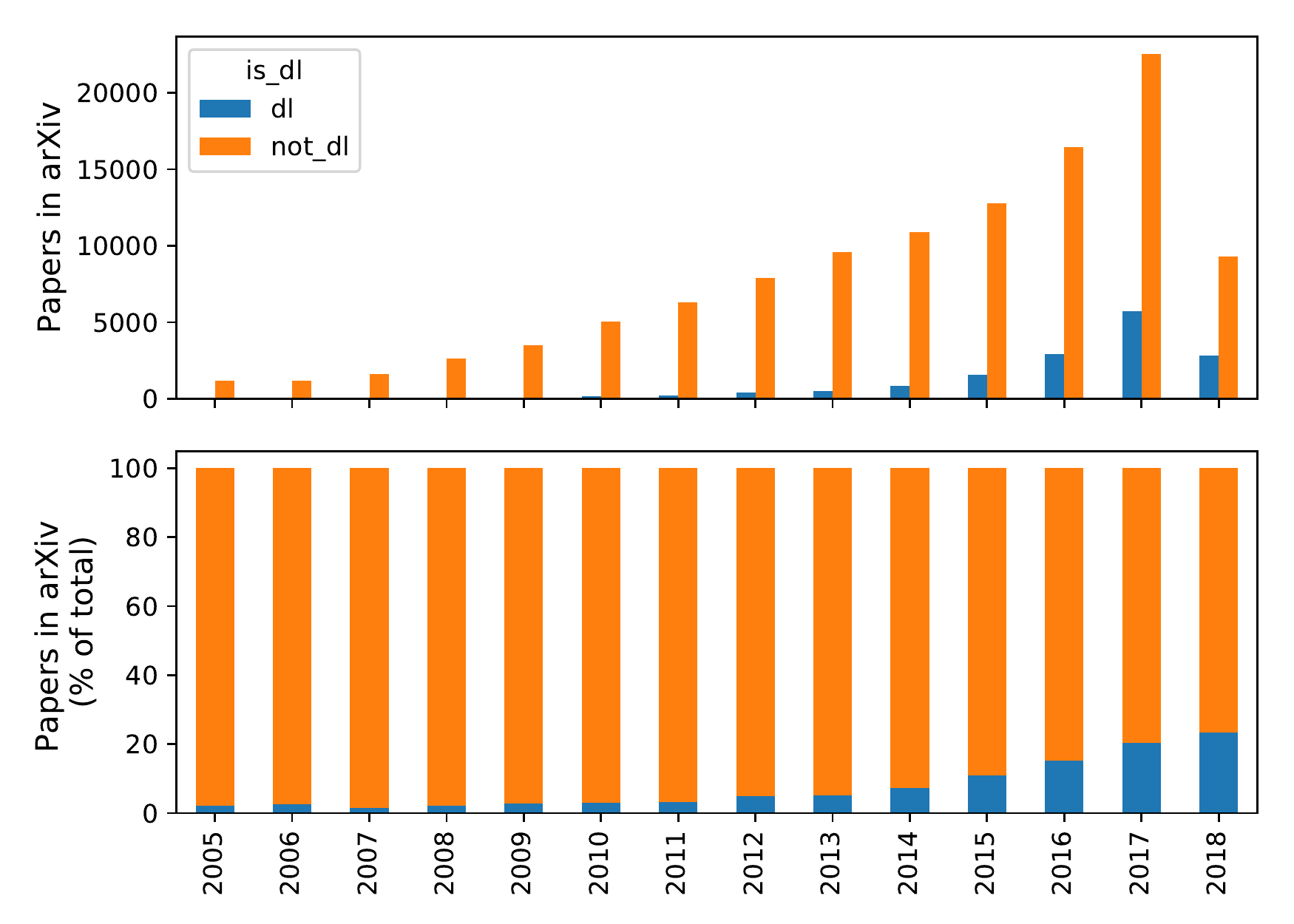}
\caption{Publication activity in \arXiv (2008-2012)}
\label{fig:total_trends}
\end{figure}

  Figure~\ref{fig:total_trends} presents the evolution of DL and non-DL publishing in \arXiv. It shows that \arXiv is becoming an increasingly popular venue for computer science research, and that DL is gaining relative importance in it. The share of DL papers in the total has grown fivefold, from 3\% before 2012, to 15\% afterwards \footnote{The results are similar if we focus on the most highly cited papers every year.}.

\subsubsection{Generality}\label{subsubsec:diffusion}
  The second GPT test for a technology is \textit{rapid diffusion in new fields}: is DL being adopted in multiple domains or restricted to a small number of areas? To assess this, we measure the number of DL papers in different \arXiv subjects {\footnote{As mentioned, most papers are labeled with multiple \arXiv subjects. We allocate a paper to a subject if it appears in it at least once.}.
  
\begin{figure}[!h]
\includegraphics[width=\textwidth]{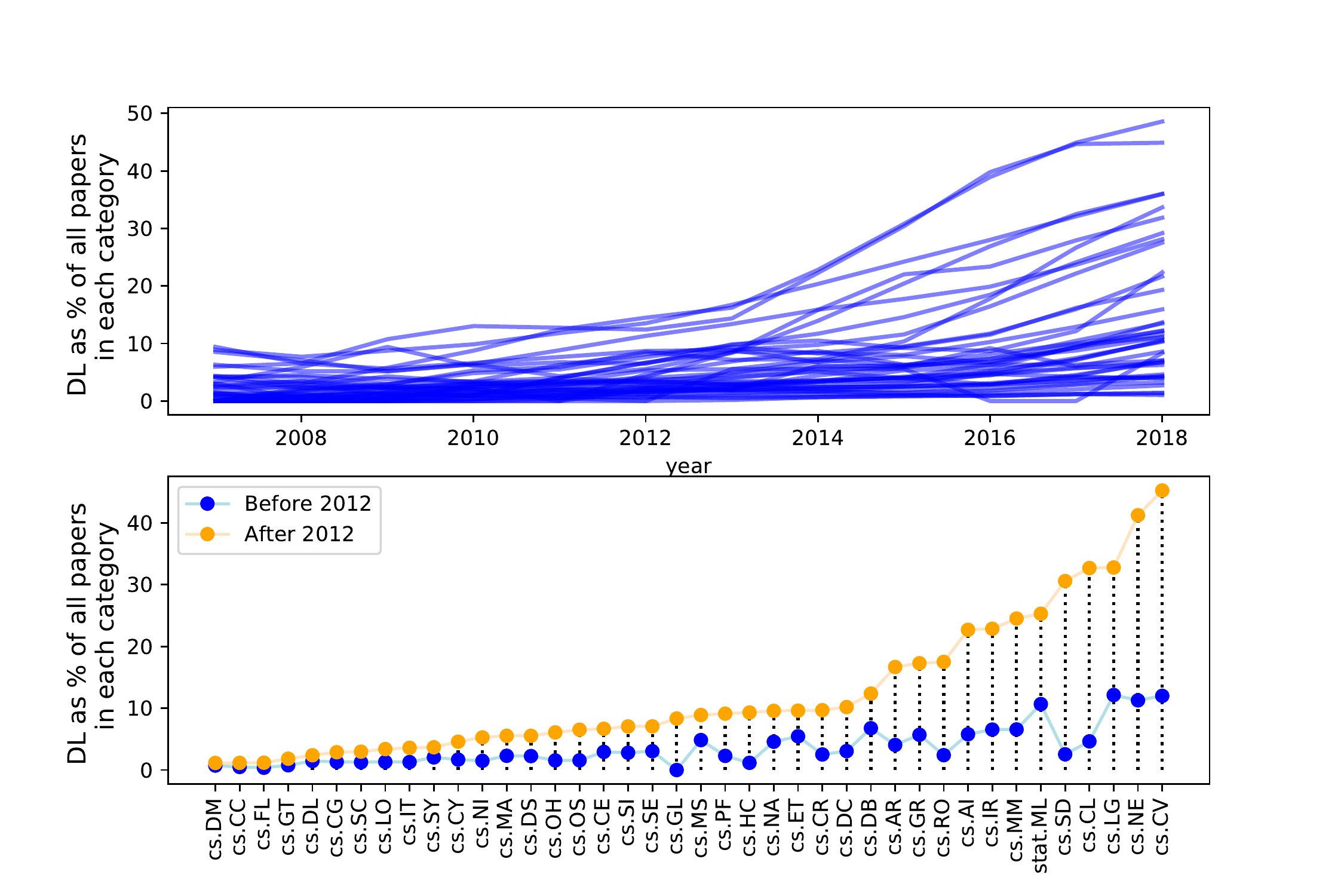}
\caption{DL as a share of activity in different \arXiv subjects. Top panel shows yearly trends for all subjects in the \arXiv data. Bottom panel compares shares of DL activity in a subject before and after 2012.}
\label{fig_9:trend_cats}
\end{figure}
  
  Figure~\ref{fig_9:trend_cats} presents the results. The top panel displays yearly changes in the shares of DL by \arXiv subject (based on 3 year moving averages), and the bottom panel compares shares of DL activity in a category before and after 2012, focusing on the top 35 computer science subjects in \arXiv by total levels of activity. 
  
  DL also fulfills the second GPT test, with a visible upward trend in the relative importance of DL in many computer science subjects, specially since 2012, the year of publication for \cite{krizhevsky2012imagenet}, a landmark paper in the use of DL in computer vision. Further, the bottom panel of \ref{fig_9:trend_cats}, shows that virtually all computer science subjects in our corpus have experienced an increase in the relative importance of DL research since 2012. As before, this is particularly visible in subjects that use unstructured data (e.g. Computer Vision) or specialize in the development of AI and ML methods (Neural Networks, Computer Learning etc.) \cite{goodfellow2016}.

\subsection{Impact in other fields}
  The third GPT test is \textit{impact in new fields}: does DL generate follow-on innovations in the fields that adopt it? Following convention, we use citations as a proxy for that impact. 
  
  Figure \ref{fig_10:DL_shares} compares the shares of DL papers in a \arXiv subjects with their share of \textit{highly cited papers} in that same subject over different periods \footnote{Highly cited papers are those in the top citation quartile for each year.}. In all cases, most \arXiv subjects are above the diagonal (this is, DL papers are overrepresented among the highly cited ones in the subject). This pattern becomes more apparent over time, supporting the idea that DL is becoming more influential in the fields where it is applied.
  
\begin{figure}[!h]
\includegraphics[width=\textwidth]{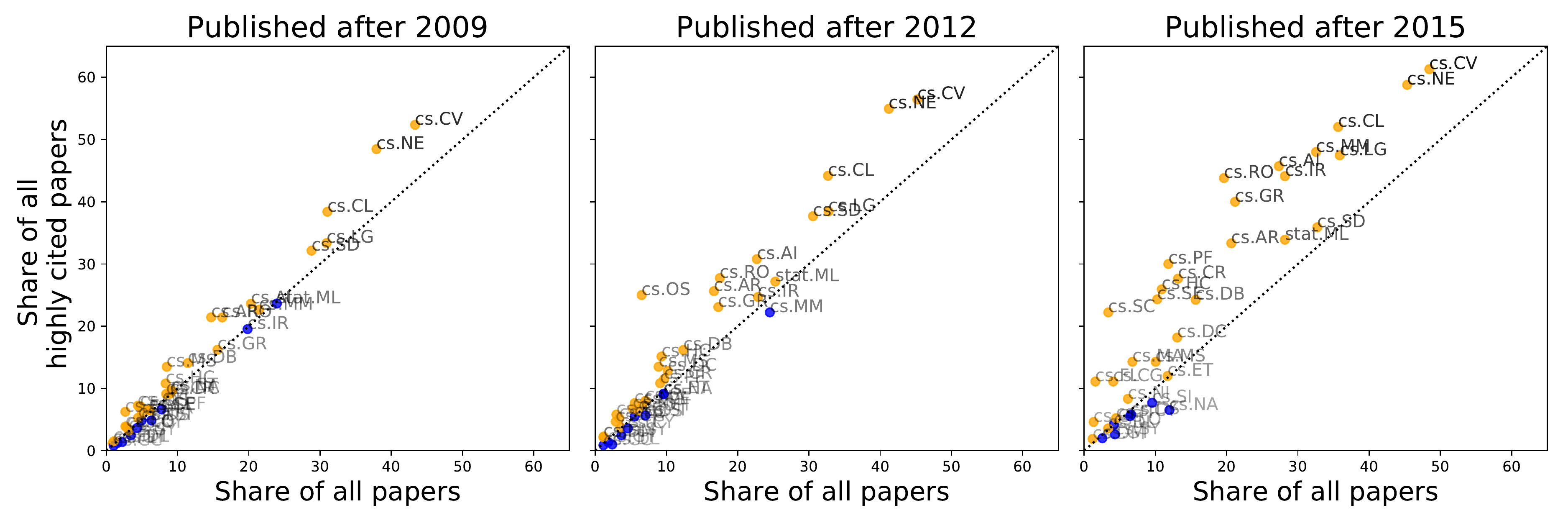}
\caption{DL papers as a share of all papers in an \arXiv category, and as a share of all highly cited papers for papers published after 2009 (left panel), 2012 (center panel) and 2015 (right panel).}
\label{fig_10:DL_shares}
\end{figure}
  
  Together, these results support the idea that DL is a GPT: its levels of activity are growing rapidly, it is spreading into more fields, and it is generating an impact (or at least attracting attention, in terms of the number of citations it receives) in the fields where it is applied.

\subsection{Evolution in the Geography of DL research}
\label{subsec:dl_geo_analysis}

  We now turn to the analysis of the geography of DL research, considering whether its evolution follows the cycle of volatility and consolidation we would expect based on the literatures reviewed in Section \ref{sec:gpt_geo}. To do this, we analyze changes in national and regional DL specialization using relative comparative advantage (RCA) indices. We define the $\text{RCA}_{dl}$ of a location (country or region) $i$ as:
  
  \begin{equation}
  \text{RCA}_{dl,i} =\frac{(\dfrac{A_{dl,i}}{A_{c,i}})}{(\dfrac{A_{dl,n}}{A_{c,n}})}
 \end{equation}
 
  Where $A_{dl,i}$ and $A_{c,i}$ are the research activity of the location in DL and in all \arXiv categories, and $A_{dl,n}$ and $A_{c,n}$ are the totals of DL activity and activity in all \arXiv categories in all locations. A $\text{RCA}_{dl,i}$ above 1 implies that the country is relatively specialized in DL, while the opposite is true if the $\text{RCA}_{dl,i}$ is below 1. $\text{RCA}$s allow us to measure changes in DL research while controlling for rapid growth in computer science activity, and for differences in size between locations. Since RCAs tend to lose robustness in observations with low levels of activity, we focus our analysis on the larger countries and regions. We also remove low quality papers from the data by focusing on those above the median of citations for the year when they were published. 
  
  Figure \ref{fig_11:country_map} presents DL specialization by country after 2012 (map in the right panel) and changes in DL specialization since 2012 for the most active countries. It shows that China has the strongest comparative advantage in DL R\&D. Interestingly, this has not changed significantly since 2012, suggesting that the development of advanced AI capabilities in China predate the recent explosion of interest in DL. We also see rapid growth in the specialization of other Asian countries such as Singapore and Korea. By contrast, all European countries in the chart with the exception of the United Kingdom and France have become less competitive in DL research (and France has, in any case, low levels of specialization in the DL). Canada and the US have also increased their DL specialization since 2012.  
  
  These changes are consistent with the idea of volatility in the early stages of GPT development, with some countries climbing up in the research rankings rapidly while others fall behind. It is also interesting to note, qualitatively, that the trends we observe echo popular narratives about the current state of the `AI race', with China in the ascendant while European countries fall behind in relative terms. After an initial slow response to the emergence of DL, the US is catching up \cite{cockburn2018impact}.  
 
\begin{figure}[!h]
\centering
\includegraphics[width=\textwidth]{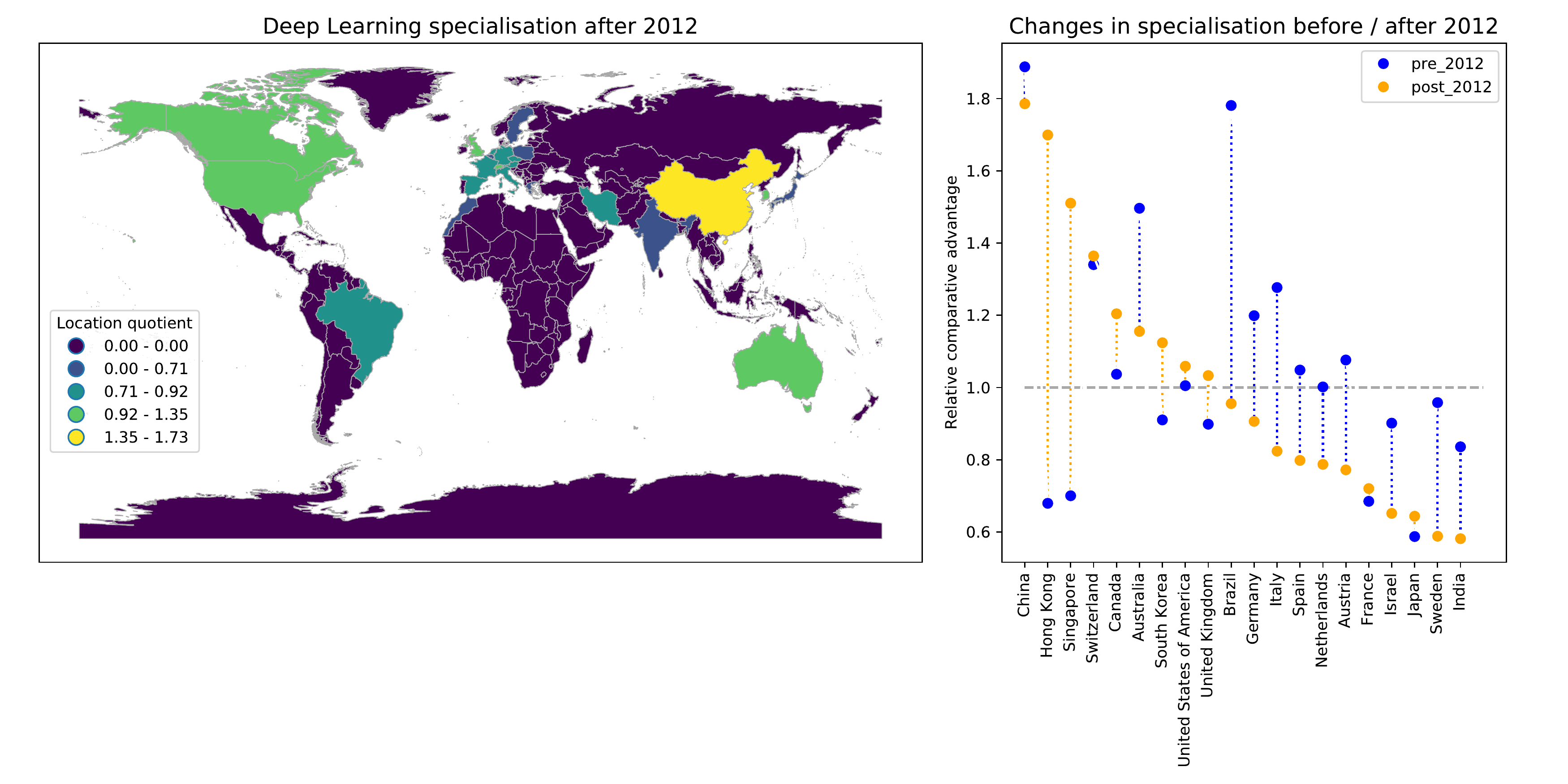}
\caption{The map in the left panel shows $\text{RCA}_dl$ by nation for papers published after 2012, focusing on papers above the median of citations in their publication year, and countries in the top 90 percentile for total level of activity. The figure in the right panel compares changes in $\text{RCA}_{dl}$ between the period before 2012 and afterwards, focusing on the top 20 countries by total level of DL activity.} 
\label{fig_11:country_map}
\end{figure}

  Figure \ref{fig_12:reg_map} presents similar figures but this time focusing on regions.  The map shows high levels of activity in a small number of regions in the East and West coast of the US, Canada, China and East Asia, Central Europe, France, Britain and Adelaide in Australia (which hosts the Australian Institute for Machine Learning Research). The right-hand panel shows US states such as Maryland, California and New York becoming more specialized in DL since 2012. Perhaps the most notable change is in Oxfordshire in the UK, which has multiplied its $\text{RCA}_{dl}$ more than seven-fold since 2012. Interestingly, we see that most of the largest regions in DL activity have also gained specialization in DL, suggesting potential advantages to scale in developing a DL research cluster. One potential explanation we explore in \ref{subsec:drivers} is that these larger regions have sufficient scale to host the combination of research and industrial capabilities required to develop the DL GPT.  

\begin{figure}[!h]
\centering
\includegraphics[width=\textwidth]{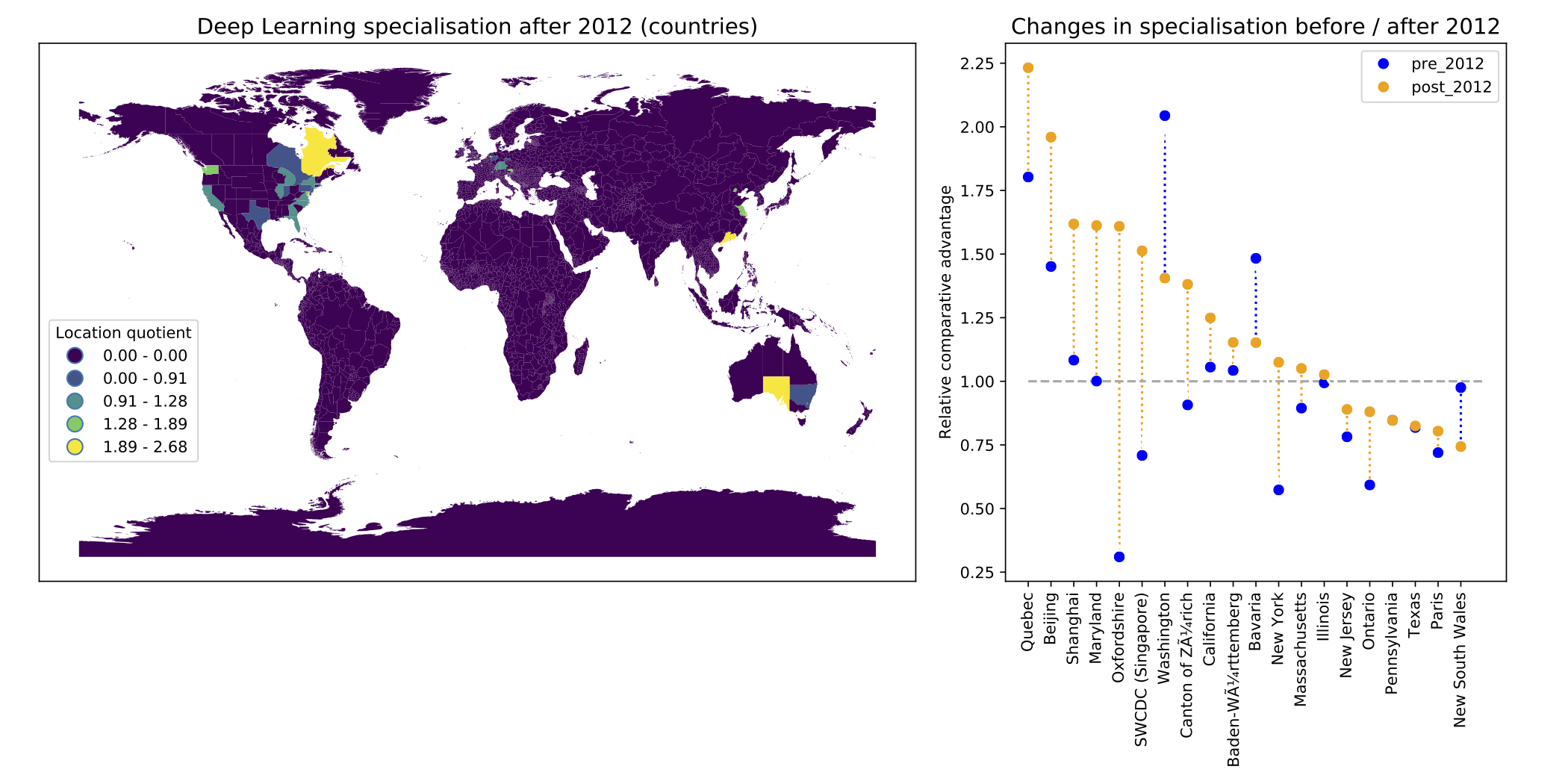}
\caption{The map in the left panel shows $\text{RCA}_{dl}$ by region for papers published after 2012, focusing on papers above the median of citations for papers in their publication year, and region in the top 99 percentile for total level of DL activity. The figure in the right panel compares changes in $\text{RCA}_{dl}$ between the period before 2012 and afterwards, focusing on the top 20 regions by total level of DL activity.} 
\label{fig_12:reg_map}
\end{figure}

  We conclude by considering changes in the dispersion and concentration of DL activity since 2009. Does the geography of DL research follow the cycle of volatility and consolidation we expect from the product life-cycle literature?
  
  Figure \ref{fig_13:vol_count} shows the recent evolution in volatility and concentration of DL activity in the largest nations and regions, focusing again on highly cited papers.

\begin{figure}[!h]
\includegraphics[width=\textwidth]{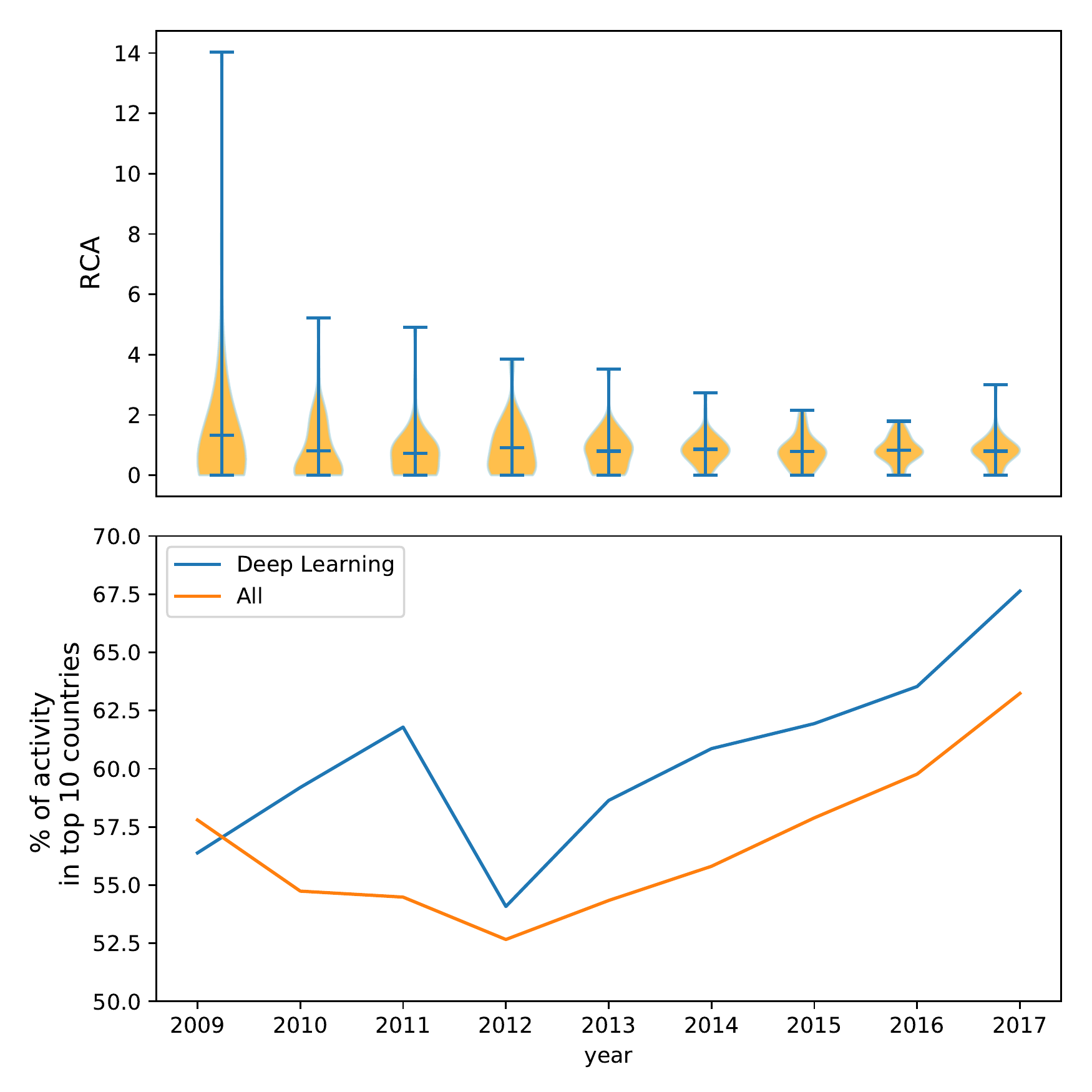}
\caption{The top panel shows the evolution in the dispersion of $\text{RCA}_{dl}$ by country between 2009 and 2017 only considering papers above the citation median for the year, and the top 50 countries by level of activity in \arXiv. The bottom panel shows the percentage of highly cited papers that concentrate in the top 10 countries}
\label{fig_13:vol_count}
\end{figure}

\begin{figure}[!h]
\includegraphics[width=\textwidth]{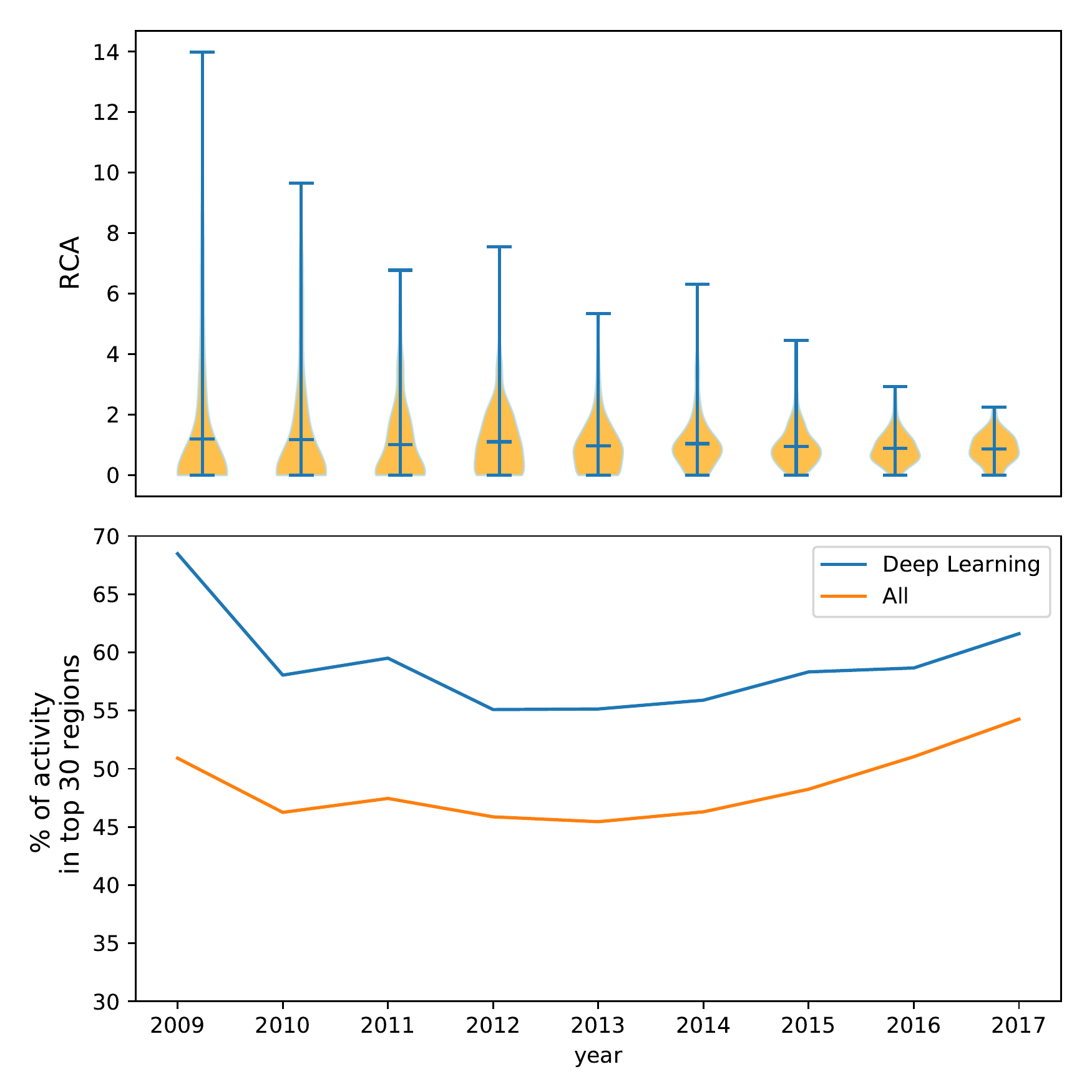}
\caption{The top panel shows the evolution in the dispersion of $\text{RCA}_{dl}$ by region between 2009 and 2017 only considering papers above the citation median for the year, and the top 150 regions by level of activity in \arXiv. The bottom panel shows the percentage of highly cited papers that concentrate in the top 30 regions}
\label{fig_14:vol_reg}
\end{figure}

  The patterns in the violin-plots in the top panel are consistent with the idea that DL experienced an initial phase of volatility (high dispersion and a flatter distribution in RCAs) followed by growing stability (less dispersion and a normal distribution with fewer locations displaying high RCAs). Also in line with what we expected, the bottom panels show a sudden decline in the shares of activity accounted for by the top countries / regions around 2012, followed by an increase in concentration afterwards. Having said this, the pattern of dis-location in DL is also present in the broader \arXiv corpus, suggesting that changes in concentration could be influenced by other factors, such as growing use of \arXiv or lower barriers to entry at the beginning of the period, with open resources such as \arXiv allowing more locations to participate in computer science research. These are all interesting questions to explore in further work. 
  
  Our analysis also shows that DL research is more geographically concentrated than computer science overall. One potential explanation is that DL research is more complex, requiring proximity for successful collaboration (\cite{balland2017geography} find something similar in their analysis of complex technologies using patent data). This is what we would expect in a GPT that relies on coordination between developers and adopters. We focus on that interaction for the remainder of this section.

\subsection{Drivers of DL cluster emergence}
\label{subsec:drivers}
  After showing that DL behaves like a GPT in its growth, diffusion, impact and geography, we turn to the analysis of the local drivers associated with its development. As we said, GPTs benefit from coordination between developers and adopters: developers aware of market needs can customize and promote their technologies to new industries. Adopters aware of GPT opportunities can find new ways to apply these technologies to their own situation. We would expect this mutual awareness to be higher when developers and adopters are close to each other, making it easier to collaborate, network and share knowledge. This means that regions where developer and adopter sectors co-locate should be more competitive in the development of a GPT.
  
  We test this hypothesis with the following model specification:
  
  \begin{equation}
  \centering
  \begin{aligned}
\text{RCA}_{dl,t1} = & \beta_0 + \beta_1\text{RCA}_{dl,t0} + \beta_2\text{arXiv}_{sp} + \beta_3\text{CrunchBase}_{sp} + \\ & \beta_4\text{arXiv}_{sp}\text{CrunchBase}_{sp} + \beta_5\text{arxiv}_{sp}*\text{CrunchBase}_{tot} + \\ 
& \beta_6\text{arXiv}_{tot} + \beta_7\times \text{is\_China} + \epsilon 
  \end{aligned}
  \label{eq:model}
  \end{equation}

  In it, we estimate the link between DL specialization after 2012 ($\text{RCA}_{dl,t1}$) and the presence of related research and industrial capabilities ($\text{arXiv}_{sp}$ and $\text{CrunchBase}_{sp}$) and their interaction ($\text{arXiv}_{sp} \times \text{CrunchBase}_{sp}$) before 2012, capturing the idea of GPT complementarities between research and industry.\footnote{The measures of related activity weight levels of regional specialization in research subjects and industrial activities by the DL similarity vectors described in \ref{sec:data} and \ref{subsec:descr}.} We also include an interaction between relevant research capabilities and total \cb activity ($\text{arXiv}_{sp} \times \text{CrunchBase}_{tot}$) to capture the benefits from deploying a GPT in industries less directly related to it. 

  We control for the levels of specialization in DL before 2012, total \arXiv activity, and a dummy for whether a region is Chinese or not ($\text{is\_China}$). We take the logarithm of all totals, calculate z-scores for all variables and focus our analysis on regions in the highest level of \arXiv activity (i.e. the top quartile) to reduce noise in the RCAs and remove a long tail of regions with little or no DL activity.\footnote{Our results are robust to changes in these thresholds}

  The correlation matrix in Figure \ref{fig_15:correlation} shows an association between $\text{DL}_{t1}$ and several independent variables and controls, including China. The correlation between between $\text{arxiv}_{sp}$ and $\text{CrunchBase}_{sp}$ is low, suggesting that locations with high specialization in research subjects relevant for DL do not always specialize in relevant industries.  Strong correlations between some independent variables suggest the presence of multicolinearity \footnote{During our robustness tests we have removed some of these interaction terms without significant changes in the results.}. 

\begin{figure}[p]
\includegraphics[width=\textwidth]{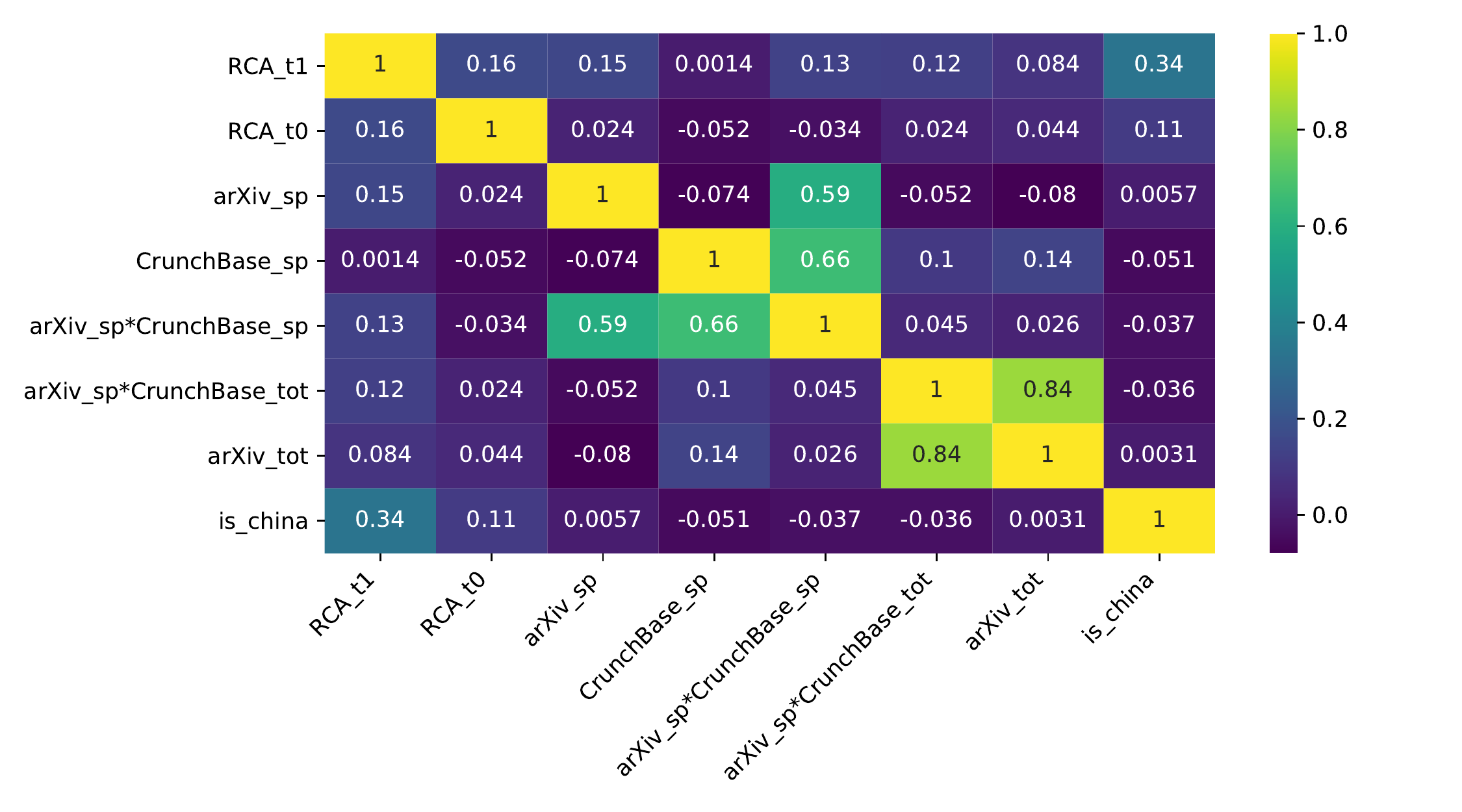}
\caption{Correlation matrix between key variables in our model. All variables have been normalized.}
\label{fig_15:correlation}
\end{figure}

Table \ref{tab_3:reg} presents the results of our regression analysis with different specifications. Model 4 is the specification in \ref{eq:model}. We review some key results:

\begin{table}[p]
\centering
\begin{tabular}{lllll}
\toprule
{} &   Model 1 &   Model 2 &   Model 3 &   Model 4 \\
\midrule
y                       &    RCA\_t1 &    RCA\_t1 &    RCA\_t1 &    RCA\_t1 \\
RCA$_{t0}$                  &   0.12*** &  0.121*** &  0.124*** &  0.126*** \\
                        &   (0.044) &   (0.044) &   (0.044) &   (0.043) \\
arXiv$_{sp}$                &  0.155*** &  0.156*** &    -0.012 &     0.006 \\
                        &   (0.044) &   (0.044) &   (0.084) &   (0.084) \\
CrunchBase$_{sp}$           &           &     0.023 &   -0.162* &    -0.135 \\
                        &           &   (0.044) &    (0.09) &    (0.09) \\
arXiv$_{sp} \times $CrunchBase$_{sp}$  &           &           &   0.261** &   0.229** \\
                        &           &           &   (0.111) &   (0.111) \\
arXiv$_{sp} \times $CrunchBase$_{tot}$ &           &           &           &   0.207** \\
                        &           &           &           &    (0.08) \\
arXiv$_{tot}$               &    0.09** &    0.086* &   0.092** &    -0.083 \\
                        &   (0.044) &   (0.044) &   (0.044) &   (0.081) \\
is\_China                &   1.54*** &  1.545*** &  1.549*** &  1.586*** \\
                        &   (0.213) &   (0.213) &   (0.212) &   (0.212) \\
$R^2$                      &     0.147 &     0.146 &     0.154 &     0.165 \\
n                       &       451 &       451 &       451 &       451 \\
\bottomrule
\end{tabular}
\caption{Dependent variable is $RCA_{dl,t1}$. Standard errors in brackets are clustered by country. *** p<0.01, ** p<0.05, * p<0.10.}
\label{tab_3:reg}
\end{table}

\begin{enumerate}
\item  There is a robust link between a region's specialization in DL before 2012 and afterwards. This suggests that the volatility in the geography of DL we described above is not absolute, with some DL specialization persisting over time.
\item There is a significant link between the interactions of $\text{arXiv}_{sp}$ with $\text{CrunchBase}_{sp}$ and with $\text{CrunchBase}_\text{tot}$, and $\text{DL}_{t1}$. This supports the hypothesis that GPT development benefits from the co-location of developers and adopters. Interestingly, once we consider this complementarity, the link between related research activity and a region's comparative advantage in DL loses significance. This suggests that the presence of relevant industries is an important ingredient in the development of a DL cluster.
\item The link between the $\text{is\_China}$ dummy and the development of a DL cluster after 2012 is strong and significant after we control for other explanatory factors such as regional research and industrial levels of activity. Together with the low $R^2$ of our models, this suggests that we our model is missing important national and regional factors that play a role in the development of DL clusters such as access to skills and data, infrastructure, regulation and supportive policies \cite{brundage2016modeling}. We plan to bring them into the analysis in future work.
\end{enumerate}

  We conclude by comparing model outputs for DL with other computer science subjects using the same specification (while focusing in the relevant research and industrial activities for each subject). One could think of these other subjects as quasi-controls allowing us to explore whether the patterns we detect in DL are also present in other fields, or DL is unique in some way. Through this, we also attempt to control for other trends which could be driving our results, such as secular changes in the usage of \arXiv. 

The results in Figure \ref{fig_16:reg_comp} shows that, in general, the interactions between \arXiv and \cb activity that we have detected in DL are not pervasive amongst other DL subjects. Interestingly, complementarities between research and industry are more important for data-related subjects such as Computer Vision, Computer Learning, Machine Learning or Computer Learning. Other subjects such as Data Structures (\texttt{cs.DS}), Network architecture ({\texttt{cs.NI}), Social and Information Networks or Logic (\texttt{cs.LO}) seem less reliant on these complementarities, perhaps because they are more mature (reducing the need for coordination between developers and adopters).\footnote{The exception to this, Information Theory (\texttt{cs.IT}) appears to be a catch-all subject present in almost 20\% of the computer science \arXiv corpus} It is also worth noting that DL and Computer Vision are the main subjects with a strong and positive association between the $\text{is\_China}$ dummy and subject specialization, suggesting that China has specific endowments that facilitate the development of these subjects, such as large unstructured datasets and targeted policies .\footnote{This result will also be driven by the overlaps between DL and Computer Vision outlined in \ref{subsec:gpt_dl_analysis}}
\begin{figure}[!h]
\includegraphics[width=\textwidth]{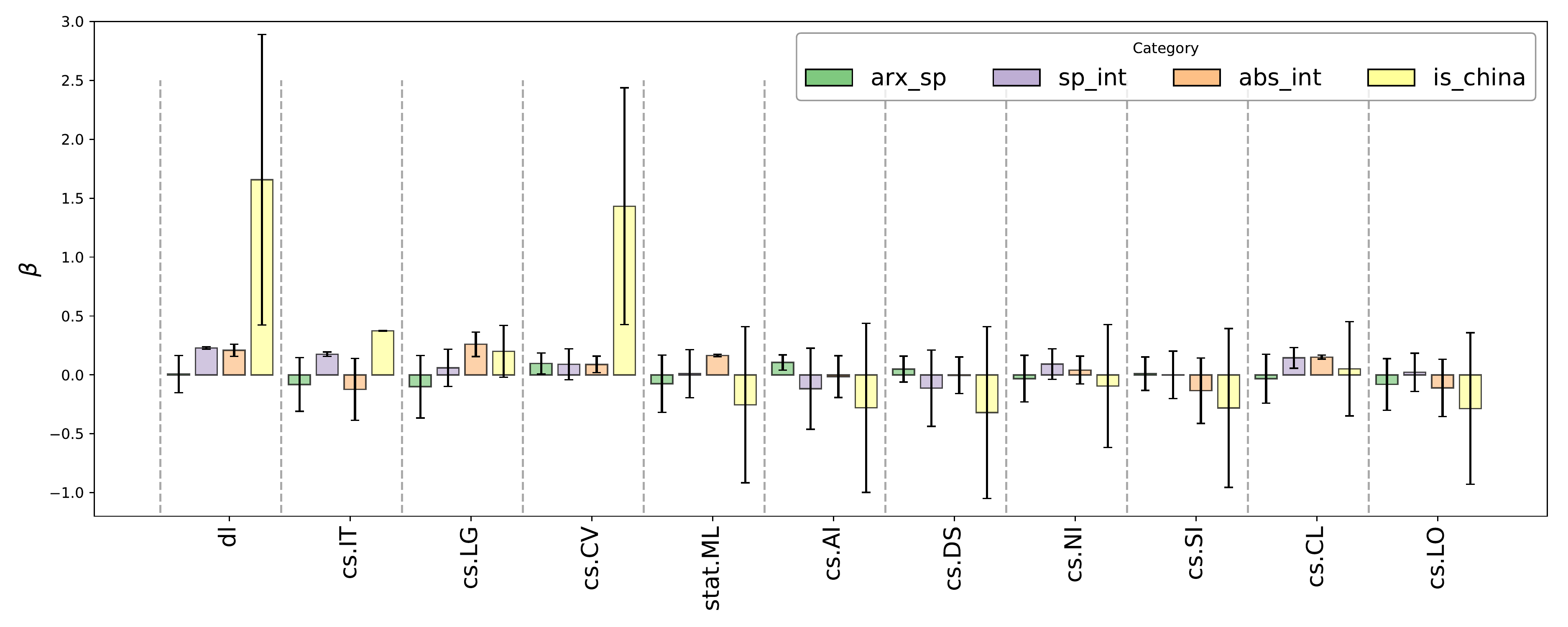}
\caption{Regression coefficients and confidence intervals for models using the specification in \ref{eq:model} in \arXiv subjects with the highest levels of activity}
\label{fig_16:reg_comp}
\end{figure}

\section{Conclusion}
\label{sec:conclusion}

\subsection{Discussion and implications}
\label{sec:discussion}

  We have studied the geography DL, a new paradigm for AI. Our analysis of \arXiv, a popular preprints website used by researchers in academia and industry supports the idea that DL has the features of a GPT technology: it has experienced rapid growth and is being applied in an increasing number of computer science subjects where it generates high-impact work (which we proxy with citations). This confirms the conclusions of previous studies such as \cite{cockburn2018impact}, and also suggests that in spite of recent criticisms of the DL paradigm, and in particular the lack of robustness stemming from its reliance on large datasets for training \cite{marcus2018deep}), researchers in multiple domains of computer science who are perhaps less likely to be swayed by hype than policymakers and entrepreneurs, are applying it in ways that their peers find interesting and useful. 

  If DL is a GPT, what are the geographical dimensions of its development? Our review of the literature suggested that the emergence of a GPT might involve an initial shift in the geography of research as new `entrants' come into the scene, followed by consolidation as central hubs of activity emerge. Our analysis at the national and regional level support this idea: we see international shifts in activity since 2012, when DL started to gain visibility, followed by growing geographical concentration. We also note that DL is at all points more geographically concentrated than computer science research, lending support to the hypothesis in \cite{goldfarb2018ai} that knowledge spillovers in AI research are localized, justifying national and sub-national policies to support its development.

  This higher geographical concentration also suggests that DL researchers benefit from co-location. We have further studied this idea with a model that estimates the link between co-location of relevant research and industrial capabilities and DL development. The results of the analysis, considering DL on its own and comparing it with other computer science subjects, supports the idea that the co-location of researchers able to develop a GPT and adopters who can explore its application favors the development of stronger DL research clusters. This result is also present in other data analytics subjects, highlighting the link between AI and broader trends towards `datafication' in the economy and society \cite{viktor2013big}.

  In terms of policy, our findings suggest that the attention that DL is attracting from national and regional policymakers is warranted by its GPT nature and evidence of localized knowledge spillovewrs. What is less clear is the extent to which the `window of opportunity' to enter the field remains open given the growing concentration of DL research activity that we identify. Our findings also echo the public narrative about the emergence of China as a global AI leader (together with the USA, Canada and Asian countries such as Singapore and Korea and, perhaps to a lesser extent, the UK), while EU countries lag behind. 

  Our analysis of the drivers of DL cluster development support the idea that co-location and collaboration in dense ecosystems of research and industrial activity offers a fertile ground for the development of GPTs that rely on new combinations of ideas from various fields and applicable in multiple sectors. Proximity between researchers and businesses could address some of the coordination failures between GPT developers and adopters identified in the literature \cite{bresnahan1995general}. One important challenge for policymakers is how to enhance these complementarities without exacerbating regional inequalities. While a geographical diversity of needs could justify dispersing research geographically so as to explore DL opportunities in a wider set of industrial and social contexts, this might weaken agglomeration economies and knowledge spillovers derived from clustering. New, detailed and timely sources of data such as those we use in this analysis can help understand and balance these trade-offs.

\subsection{Limitations and issues for further research}
\label{sec:limitations}
  Our use of \arXiv data raises some concerns. To begin with, this is a platform with low barriers to entry, so many of the papers there might be of low quality. We have tried to address this problem by matching the \arXiv data with \MAG, and focusing key parts of our analysis on highly cited, hopefully higher quality papers. Future work should expand and further validate our conclusions in other data sources such as patents or open source projects. 

  Second, to which extent does our research data capture changes in technology development and business diffusion? Throughout our analysis we have assumed that the clustering of DL research is a good proxy for DL R\&D development activities with an industrial application. Although anecdotal evidence suggests high level of industry participation in \arXiv, and we find a strong correlation between the levels of activity in \arXiv and \cb, there is risk of biases if different research communities, sectors or countries display variation in their propensity to publish their work in \arXiv. Further triangulation of \arXiv data with other sources, including peer-reviewed research in comparable disciplines, as well as industry patenting and the financial performance of companies in DL-related sectors, would help to address these concerns.
  
  Third, there is the issue of causality. While our analysis has a longitudinal dimension, and qualitatively controls for unobservables by comparing DL model estimates with other computer science subjects, we cannot rule out that other local factors such as access to skills and finance or a supportive policy environment might be underpinning the links between research and industrial activity and DL research clustering that we have detected. Going forward, we would like to incorporate in our analysis shocks to industrial activity with an exogenous element, such as regulatory changes, or industrial policy interventions so as to identify more precisely the causal effects of research/industry co-location in DL cluster development. 

There are many interesting directions to extend our work:

  First, our analysis says little about the mechanisms behind the link between research / industry co-location and DL cluster development: are these links driven by knowledge spillovers, the formation of a technical talent pool that researchers and industry both tap on, or access to finance (e.g. adopters fund development activities in regional research institutions)? A better understanding of those mechanisms would help to address the issues of causality above, and yield policy-relevant implications about what programs to put in place to strengthen DL clusters. 

  Second (and relatedly), our analysis takes a siloed view of DL research clusters, only considering geographical proximity to other DL researchers and technology businesses as a source of valuable knowledge about new techniques and business applications. In reality, researchers access this knowledge through many other channels and further afield, including via popular international collaborations such as NIPS. Going forward, we will address this by studying the network of co-authorships and citations in our data, and trying to understand the role of international conferences in the dissemination of knowledge in DL. This analysis could reveal cross-country flows of ideas and collaborations going against the narrative of a zero-sum global AI race dominating popular debates.
  
  Third and last, we have not considered in detail the technological characteristics of the DL `dominant design': what are its features and components, and how stable are they? What are the parallel paths for DL that have been explored and set aside? Should some of them be maintained to avoid a premature lock-in to suboptimal standards for the large-scale deployment of the AI GPT \cite{aghion2009science}}? As we mentioned before, some researchers have expressed concerns about the lack of robustness and interpretability in DL systems, calling for their combination with older paradigms for AI development. New techniques and methods are being developed in response to this. Identifying what they are, and overlaying their geography with the geography of DL explored in this paper could yield a richer understanding of the diversity of evolutionary paths for emerging technologies, and their spatial dimensions.  Rich text data from papers could be marshaled for this, using the same NLP approach we followed in this paper. 
 
 All these ideas highlight the analytical and policy opportunities for using new data sources for the analysis of emerging technologies, and turning  AI-related methods and tools towards the analysis of AI itself.
  
\bibliographystyle{unsrt}
\bibliography{sample}

\end{document}